\documentstyle[aps,prl,preprint,floats,epsfig]{revtex}

\long\def\simplex#1#2#3#4{
\begin{figure}[#1]	
   \begin{center}
   \quad
   \quad
   \vbox{
   \quad 
   \hskip +10.0cm\parbox[t]{16.5cm}{\psfig{figure=#2,width=15.5cm} 
   \vskip -0.cm
   \caption[]{\small  \label{fig:#3} #4 }}
   }
   \vskip -0.5cm
   \quad
   \end{center} 
\end{figure}
}

\def\piz{\pi^0}

\def\BR{{\cal B}}

\def\thrhp0{$\tau^-\to h^-h^+h^-\piz\nu_{\tau}$}
\def\lsim{\stackrel{<}{\sim}}

\begin{document}

\preprint{\tighten\vbox{\hbox{\hfil CLNS 99/1631}
                        \hbox{\hfil CLEO 99-11}
}}

\title{Resonant structure of $\tau\to 3\pi\piz\nu_{\tau}$ and 
$\tau\to \omega\pi\nu_{\tau}$ decays}  

\author{CLEO Collaboration}
\date{\today}

\maketitle
\tighten

\begin{abstract} 
The resonant structure of the four pion final state in the decay
$\tau \to 3\pi\pi^0\nu_\tau$
is analyzed using 
$4.27$ million $\tau^+\tau^-$ pairs 
collected by the CLEO II experiment. 
We search for second class currents in the decay 
$\tau \to \omega\pi\nu_\tau$ 
using spin-parity analysis and 
establish an upper limit on the non-vector current contribution. 
The mass and width of the 
$\rho'$ 
resonance are extracted from a fit to the 
$\tau \to \omega\pi\nu_\tau$ 
spectral function. 
A partial wave analysis of the resonant structure of the 
$\tau \to 3\pi\pi^0\nu_\tau$ 
decay is performed;
the spectral decomposition of the four pion system 
is dominated by the 
$\omega\pi$ and $a_1 \pi$ 
final states.
\end{abstract}
\newpage

{
\renewcommand{\thefootnote}{\fnsymbol{footnote}}

\begin{center}
K.~W.~Edwards,$^{1}$
R.~Janicek,$^{2}$ P.~M.~Patel,$^{2}$
A.~J.~Sadoff,$^{3}$
R.~Ammar,$^{4}$ P.~Baringer,$^{4}$ A.~Bean,$^{4}$
D.~Besson,$^{4}$ R.~Davis,$^{4}$ S.~Kotov,$^{4}$
% ...
\end{center}
 
\small
\begin{center}
$^{1}${Carleton University, Ottawa, Ontario, Canada K1S 5B6 \\
and the Institute of Particle Physics, Canada}\\
$^{2}${McGill University, Montr\'eal, Qu\'ebec, Canada H3A 2T8 \\
and the Institute of Particle Physics, Canada}\\
$^{3}${Ithaca College, Ithaca, New York 14850}\\
$^{4}${University of Kansas, Lawrence, Kansas 66045}\\
% ...
\end{center}

\setcounter{footnote}{0}
}
\newpage

\section{Introduction}

The resonant structure of the decay $\tau \to 4\pi\nu_\tau$
has been the subject of experimental studies for the past several decades.
The first observations of $\tau\to\omega\pi\nu_\tau$ were reported by
the ARGUS~\cite{argus_ompi} and 
the CLEO~\cite{cleo_first} Collaborations in 1987. 
Since then, the ARGUS Collaboration
estimated~\cite{argus_rhopipi} the $\rho\pi\pi$
branching fractions and the ALEPH Collaboration provided improved
measurements~\cite{aleph} of
branching fractions of the $\tau\to4\pi\nu_\tau$ subchannels. No
attempt has been made, so far, to determine a complete resonant 
structure of
the four pion final state. 



In the Standard Model, $\tau$ decays to four pions 
proceed via vector current producing hadrons 
in the $J^P=1^-$ spin-parity state. At low energies,
the hadronic current is dominated by vector 
$\rho$, $\rho'$ and  $\rho''$ resonance form factors
which can be approximated 
by relativistic Breit-Wigner functions. 
The full formalism is discussed in 
Refs.~\cite{santamaria,kuhn,mirkes,li}.
The parameters of  $J^P=1^-$ resonances are of particular interest 
for phenomenological tests of the Conserved Vector Current
(CVC) hypothesis that compare multipion final states in
$\tau$ decays and $e^+e^-$ annihilations~\cite{eidelman}.
So far, there have been only two published
attempts~\cite{aleph_2pi,tau96} to extract the mass 
and width of the $\rho'$ resonance
from analyses of the $\tau\to\pi\piz\nu_\tau$ decay
and none from $\tau\to 4\pi\nu_\tau$. 

In this study, we determine the resonant decomposition of the decay
$\tau\to\omega\pi\nu_\tau$ and search for second class currents via
a spin parity analysis. We fit the 
$\tau\to\omega\pi\nu_\tau$
spectral function to obtain the relative contributions of the $\rho$,
$\rho'$ and $\rho''$ resonances,
as well as the mass and width of the $\rho'$ meson.
We assume that in a first
approximation the widths of the $\rho$, $\rho'$ and $\rho''$ resonances
do not depend on the invariant mass of four pions.
We  use the extracted values
of the mass and width of the $\rho'$ to perform a full unbinned maximum 
likelihood 
fit of the $2\pi^-\pi^+\piz$ final state.

To perform this
fit, we use several phenomenological models. Besides the $\omega\pi$
channel, we allow for $a_1\pi$, $\sigma\rho$, $f_0\rho$,
non-resonant $\rho\pi\pi$ and non-resonant $3\pi\piz$ channels, as
well as for their various combinations, and fit for their relative
contributions to the $3\pi\piz$ resonant structure. This analysis
shows clear dominance of the $\omega\pi$ and $a_1\pi$ components.

With a better knowledge 
of the four pion subchannel structure,
we then perform a fit to the $\tau\to\omega\pi\nu_\tau$
spectral function assuming a mass-dependent width for the three
$\rho$ resonances. 


We do not use particle identification to separate pions from kaons. 
A small component containing kaons (e.g., $K3\pi$) is included in 
the $\tau$ background, 
but its contribution is negligible. 
Generalization to charge conjugated reactions and 
final states is implied throughout this paper.

\section{Experiment and data selection}

We use data from the reaction $e^+e^-\to\tau^+\tau^-$  
collected by the CLEO~II experiment
at the Cornell Electron Storage Ring (CESR) at or near 
the energy of
the $\Upsilon(4S)$. The data correspond to a total 
integrated luminosity of 
$4.68\ {\mbox{fb}}^{-1}$ and contain about $4.27$ million $\tau^+\tau^-$ 
pairs. 

CLEO~II is a general-purpose solenoidal magnet 
detector~\cite{kubota}.
The momenta
of the charged particles are measured by 
a 67-layer drift chamber tracking system operating inside a 1.5~T
superconducting 
solenoid. Photons and electrons are detected in a 7800-crystal CsI
electromagnetic calorimeter. Muons are identified using 
proportional
counters placed at various depths in the return iron of the
magnet yoke.

The event selection is designed to use the kinematical properties and the low
multiplicity of $\tau$ decays to separate them from events 
copiously
produced in two-photon interactions, the process $e^+e^-\to q\bar{q}$,
Bhabha scattering, and muon-pair production.
We select events with four charged tracks in a
1-vs-3 topology by requiring that one
of the tracks must be at least 90$^\circ$ away from all the others.
When there is more than one combination
satisfying these criteria, we minimize the 
acolinearity angle between the momentum of the single track
and the total momentum of the 3-prong system
to select
the most back-to-back combination. 
The single track hemisphere is then used as a tag associated with a 
one-prong $\tau$ decay to 
$e\bar{\nu_e}\nu_\tau$, $\mu\bar{\nu_\mu}\nu_\tau$, $\pi\nu_\tau$ or
$\rho\nu_\tau$.
The other three tracks and at least two well identified photons, 
separated by more than $90^\circ$ from the tag track, are used
to reconstruct the hadronic part of the signal decay 
$\tau\to 3\pi\piz\nu_\tau\to 3\pi\gamma\gamma\nu_\tau$.  

To ensure well modeled acceptance, momenta of the tracks from 
signal and tag 
hemispheres must exceed $0.025E_{bm}$ and $0.05E_{bm}$
respectively, where $E_{bm}$ is the energy of the colliding 
electron or positron beam and typically lies between 5.26 and 5.29~GeV.
To suppress decay channels with $K_s^0\to\pi^+\pi^-$ decays, we require 
that the impact parameter of charged tracks with respect to the 
beam axis
must be less than 5~mm. The momentum of a $\piz$ candidate is
reconstructed from the energies and directions of two photons,
constrained to the mass of the $\piz$. The two photon candidates must
produce showers in the barrel part of the crystal calorimeter (i.e.,
$|\cos \theta_\gamma | <0.71$, 
where $\theta_\gamma$ is an angle between the photon and $e^+e^-$ beam
axis) and be separated by more than $20^{\circ}$ from the
closest charged track. Energy deposition in the calorimeter is
required to have a photon-like lateral profile and
be greater than 75~MeV for the photon with the smaller energy
and greater than 120~MeV for the photon with the higher energy. 
We consider only
those two-photon combinations that are within 2.5
standard deviations from the nominal $\piz$ mass. To increase 
statistics
for the rho tag, we also use showers in the tag hemisphere,
detected both in the barrel and endcaps, with
energy deposition greater than 30~MeV.

For electron identification, we require that the momentum of the 
track
be greater than 0.5~GeV/c and that
the energy deposited in
the calorimeter be consistent with the momentum of the electron
candidate measured by the drift chamber: $|(E/p-1.0)|<0.1$. We also
require that the rate of energy loss due to ionization in the drift
chamber must be greater than the expected rate minus two standard
deviations. A track is identified as a muon if it has correlated 
drift
and muon chamber hits and traverses at least three absorption 
lengths
of the material, when the momentum of the track is below 2~GeV/c, 
and at
least five absorption lengths, when the momentum of the track is 
above
2~GeV/c. 
To minimize the probability of pion-lepton misidentification for 
both electron and muon tags, we do not allow additional 
showers in the calorimeter with energy deposition greater 
than 120~MeV that are unmatched to any of the charged tracks. 
A tag track is identified as a pion if it was not identified as a 
lepton and if the invariant mass of the tag hemisphere is less than
0.5~$\mbox{GeV/c}^2$, where the invariant mass 
of the tag hemisphere is defined
as the invariant mass of the tag track, assuming the pion mass, and all
neutral showers in the tag hemisphere. A tag track is identified as 
a rho if it was not identified as a lepton, if the invariant mass of 
the two most energetic showers in the tag hemisphere is consistent with
the $\piz$ mass within 2.5 standard deviations and, finally, if the
effective mass of the tag track, assuming the pion mass, and of the 
$\pi^0$ candidate    
lies between 0.5 and 1.2~$\mbox{GeV/c}^2$. To 
reduce
combinatorial and hadronic background for the rho tag, we do not 
allow
any showers with energy deposition greater than  120 (350)~MeV in 
the
tag hemisphere if the center of the shower is more (less) than 
30~cm
from the tag track projection onto the calorimeter.

A typical event, in which the decay
$\tau\to 3\pi\piz\nu_\tau$ is tagged by a one-prong $\tau$ decay,
does not deposit any significant extra energy in the calorimeter, 
other than that
associated with either the $\piz\to \gamma\gamma$ decay or with 
interactions
between charged hadrons and CsI crystals. In most cases, additional
unmatched energy deposited in the calorimeter  
is a signature of various background events having one or more 
additional 
$\piz$'s.
To suppress these backgrounds,
we use the value of the unmatched energy to veto such 
events. We reject events if there is at least one
unmatched energy cluster in the calorimeter, which is not used for
$\piz$ reconstruction, which is more (less) than 30~cm away from 
the
closest projection of the charged track onto the calorimeter
and has energy deposition more than 75 (350)~MeV.

Further background reduction is based on the fact that, 
in a $\tau$ decay, there is always at least one neutrino carrying 
away
undetected energy and momentum. Thus, a
typical $\tau$ event has a large missing transverse 
momentum, $P_{t}$, whereas a typical two photon event has a small
missing transverse momentum and small visible energy, $E_{tot}$.
The transverse momentum and visible energy are measured using the 
four reconstructed charged tracks, assuming the pion masses for
the three charged tracks in the signal hemisphere,
and one $\piz$ (two $\piz$'s for the 
rho tag). To remove two-photon background, we use the
following selection criteria~\cite{rhovsrho}:
$P_t>0.1(2E_{bm}-E_{tot})$,\ $P_t>0.075E_{bm}$ and 
$E_{tot}>0.6E_{bm}$.
To reduce background produced by Dalitz decays $\piz\to e^+
e^-\gamma$ or by $\gamma$ conversions in the
detector, we reject any event that has a well identified electron 
among
the three charged
tracks in the signal hemisphere. To reduce combinatorial $q\bar{q}$
background, 
we impose a requirement on the undetected neutrino mass.
The neutrino mass can be calculated as:
\begin{equation}
\label{eq:m_nu}
M_{\nu}^2 = (P_{\tau}-P_q)^2 =
M_{\tau}^2 + q^2 + 
2|\vec{P_{\tau}}||\vec{P_q}|\cos\theta_{\tau q} - 2E_{bm}E_q\ .
\end{equation}
Because we cannot measure $\cos\theta_{\tau q}$, we assume it to be 1
and require that the resulting maximum value of the 
square of the undetected neutrino mass
be positive: $M_{\nu,max}^2>0$.
In Eqn.~(\ref{eq:m_nu}) the value of $M_{\tau}$ is taken as 1.777
GeV/$c^2$,
the momentum of the $\tau$ is expressed as
$|\vec{P_{\tau}}| = \sqrt{E_{bm}^2-M_{\tau}^2}$,
the symbols $q$ and $E_q$ represent the invariant mass and energy
of the four pion final state, and the energy of the $\tau$ 
is set equal to the beam energy $E_{bm}$.

The distribution of the $3\pi\piz$ effective mass, 
before the cut on the $\tau$ neutrino mass is applied,
is plotted in Fig.~\ref{fig:mccc_anu}.  
In the vicinity of the $\tau$ mass the production of $3\pi\piz$ events
is strongly suppressed by phase space and the signal-to-background 
ratio becomes small. Hence,
we accept only those events that have the effective mass of the
$3\pi\piz$ final state lying in the range from 0.9 to
1.7~$\mbox{GeV/c}^2$.

After all signal selection criteria are applied, we find $25,374$ 
data events with an estimated background of $577\pm17$ 
events coming from $e^+e^-\to q\bar{q}$ production 
and $669\pm13$ events from other $\tau$ decays.
The background estimates are based on Monte Carlo.
The latter background source includes contributions from 
$\tau^-\to2\pi^-\pi^+2\pi^0\nu_{\tau}$ and
$\tau^-\to\pi^-\pi^0 K_s\nu_{\tau} \to 2\pi^-\pi^+\pi^0\nu_{\tau}$ 
decays whose yields are estimated as 1.6\% and 0.7\%, respectively.
The $\tau$ background also includes small $K\pi\pi\piz$ and $KK\pi\piz$
components. Their contributions, calculated using the 
values~\cite{PDG} of  
$\BR(\tau^-\to K^-\pi^+\pi^-\pi^0\nu_{\tau}) = (0.08\pm0.04)\%$ 
and $\BR(\tau^-\to K^-K^+\pi^-\pi^0\nu_{\tau}) = (0.069\pm0.030)\%$
for the branching fractions,
are found to be negligible, as well as background
contamination from other $\tau$ decays.
The combined efficiency for all tags is estimated to be 6.75\%.
The measured branching fraction 
is $\BR(\tau\to3\pi\piz\nu_\tau)=(4.19\pm0.10)\%$,
which does not include a conservatively estimated 5\%\ 
systematic error associated with the data analysis. This is
is in good agreement with the Particle Data Group~\cite{PDG} 
value of $(4.35\pm 0.10)\%$.
Contamination due to $2\gamma$ physics
is estimated to be completely negligible.  

\simplex{htbp}{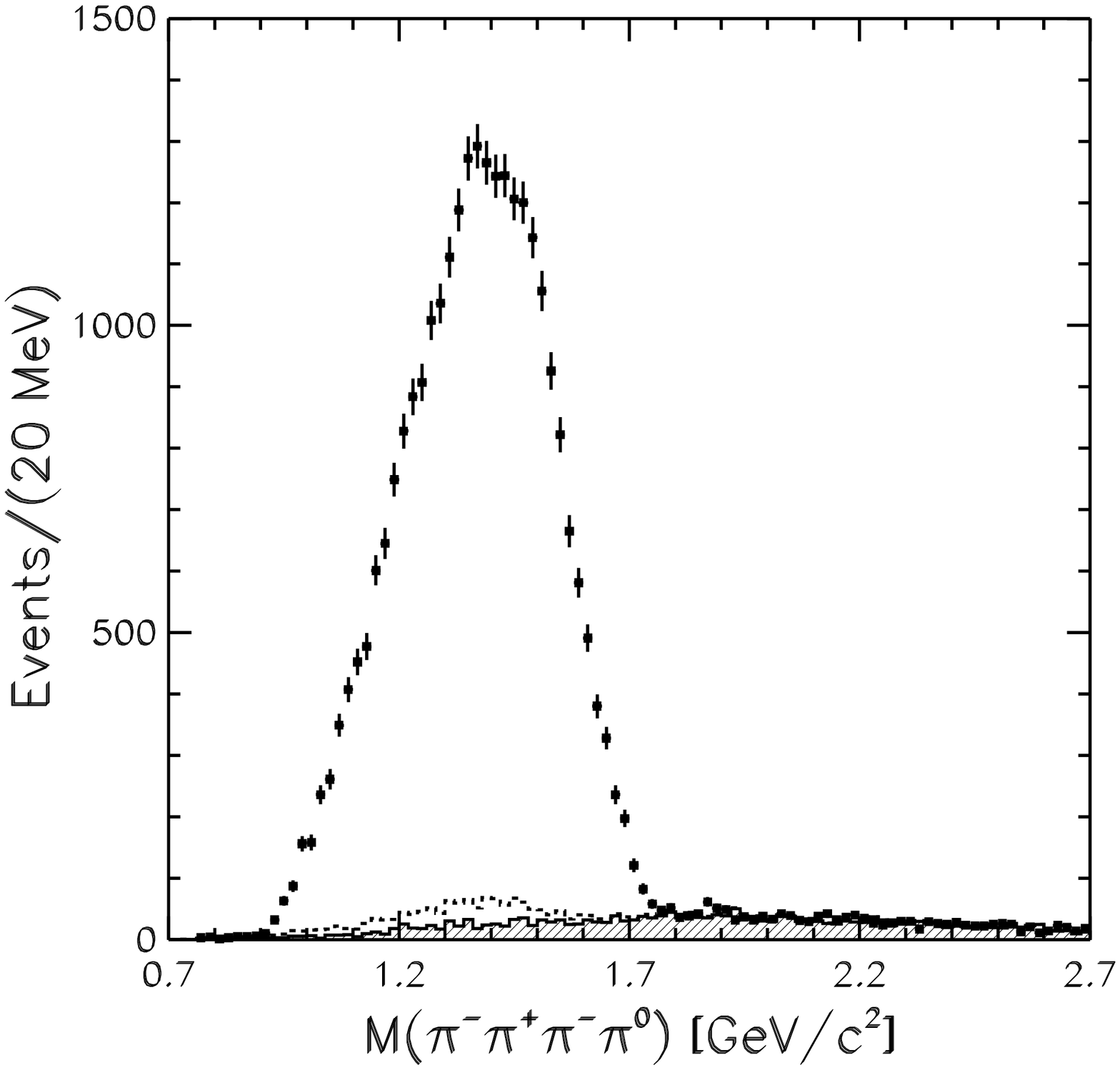}{mccc_anu}
{
The distribution of the invariant mass of $3\pi\piz$ is plotted for the 
data (crosses), 
total background (dashed line) and $q\bar{q}$ background (hatched area),
before the cut on the $\tau$ neutrino mass. 
Further analysis is limited to events between 0.9 and 1.7 $\mbox{GeV}/c^2$.
}

\section{The spectral functions in 
   $\tau\to 3\pi\piz\nu_{\tau}$ and $\tau\to\omega\pi\nu_{\tau}$
   decays}
\label{sec:spectr}

The shape of the $3\pi\piz$ effective mass in 
$\tau$ decays is distorted by phase space and helicity
factors. It is useful to examine the spectral
function independent of such effects specific to $\tau$ decays.

The spectral function for  $\tau\to3\pi\piz\nu_{\tau}$
is defined~\cite{tsai} as:
\begin{equation}
V^{3\pi\piz}(q) = \frac{d\Gamma_{4\pi\nu}(q)}{dq}
	\frac{16\pi^2M_\tau^3}{G_F^2 V_{ud}^2}
	\frac{1}{q(M_{\tau}^2-q^2)^2(M_{\tau}^2+2q^2)}\ ,
\end{equation}
where $q=M(3\pi\piz)$, $V_{ud}$ is an element of the CKM matrix, 
$M_{\tau}=1.777$~GeV/c$^2$ and 
$G_F$ is the Fermi constant. The differential partial
width can be represented as:
\begin{equation}
\frac{d\Gamma_{4\pi\nu}(q)}{dq} = \frac{1}{N} \frac{dN(q)}{dq}
	\frac{G_F^2M_\tau^5}{192\pi^3}
	\frac{\BR(\tau\to3\pi\piz\nu_\tau)}{
	\BR(\tau\to e\bar{\nu_e}\nu_\tau)}\ .
\end{equation}
Thus, the spectral function is given by:
\begin{equation}
\label{eq:spec_fn}
V^{3\pi\piz}(q)=\frac{1}{N} \frac{dN(q)}{dq} 
\frac{1}{q(M_{\tau}^2-q^2)^2(M_{\tau}^2+2q^2)}
\frac{\BR(\tau\to3\pi\piz\nu_{\tau})}{
\BR(\tau\to e\bar{\nu_{e}}\nu_{\tau})}
\frac{M_{\tau}^8}{12\pi V_{ud}^2}\ .
\label{eqn:specdef}
\end{equation}
To reduce uncertainty caused by the finite bin width,
the factor $q(M_{\tau}^2-q^2)^2(M_{\tau}^2+2q^2)$ in the 
denominator
is averaged over the bin width. 

In full analogy with $V^{3\pi\piz}(q)$, we can define a spectral
function for the
$\tau\to\omega\pi\nu_{\tau}$ decay.  To extract the $\omega\pi$ component 
from the $3\pi\piz$ final state, we use a technique similar to the
usual sideband subtraction. For each bin of 
$q=M(3\pi\piz)$,
we select events in the $\omega$ signal region ($0.76\ 
\mbox{GeV}/c^2
<M(\pi^-\pi^+\piz)<0.81\ \mbox{GeV}/c^2$) 
and fit the combined sidebands ($0.83\ 
\mbox{GeV}/c^2<M(\pi^-\pi^+\piz)<0.90\
\mbox{GeV}/c^2$) and  
($0.60\ \mbox{GeV}/c^2<M(\pi^-\pi^+\piz)<0.74\ \mbox{GeV}/c^2$)
to a second order polynomial. The fit parameters are then used to 
estimate
the background under the $\omega$ peak. To estimate the statistical 
uncertainty
of the non-$\omega$ component in the signal region, we 
propagate
errors using a full covariance matrix. 
In addition to the statistical error for each bin, there is an  
uncertainty 
due the sideband definition. To estimate the size of this 
error for each bin, we vary the sideband definition and
change the order of the polynomial from 2 to 1.
 
In Fig.~\ref{fig:omegapi_spectral_bkg_variations}
the shape of the $M(\omega\pi)$ distribution, which includes both
$\omega$ signal and combinatorial background, is shown with solid
squares. The non-$\omega\pi$
contribution is shown with dashed and dotted lines for different
sideband definitions, and its average value is shown with open 
squares.
For every bin, the error is calculated as a sum in 
quadrature of both statistical and systematic uncertainties. 

Beginning with the distributions shown in 
Figs.~\ref{fig:mccc_anu}
and \ref{fig:omegapi_spectral_bkg_variations},
we subtract backgrounds, and apply 
efficiency corrections that were determined using a detailed Monte Carlo
simulation of the $\tau$ decay process and the detector response.
We then employ Eqn.~(\ref{eqn:specdef}) to extract
the spectral functions
$V^{3\pi\piz}(q)$ and $V^{\omega\pi}(q)$, as well as
the non-$\omega\pi$ contribution
$V^{non-\omega\pi}(q) \equiv V^{3\pi\piz}(q)-V^{\omega\pi}(q)$.
All three spectral functions are shown in 
Fig.~\ref{fig:omegapi_spectral}.

\simplex{htbp}{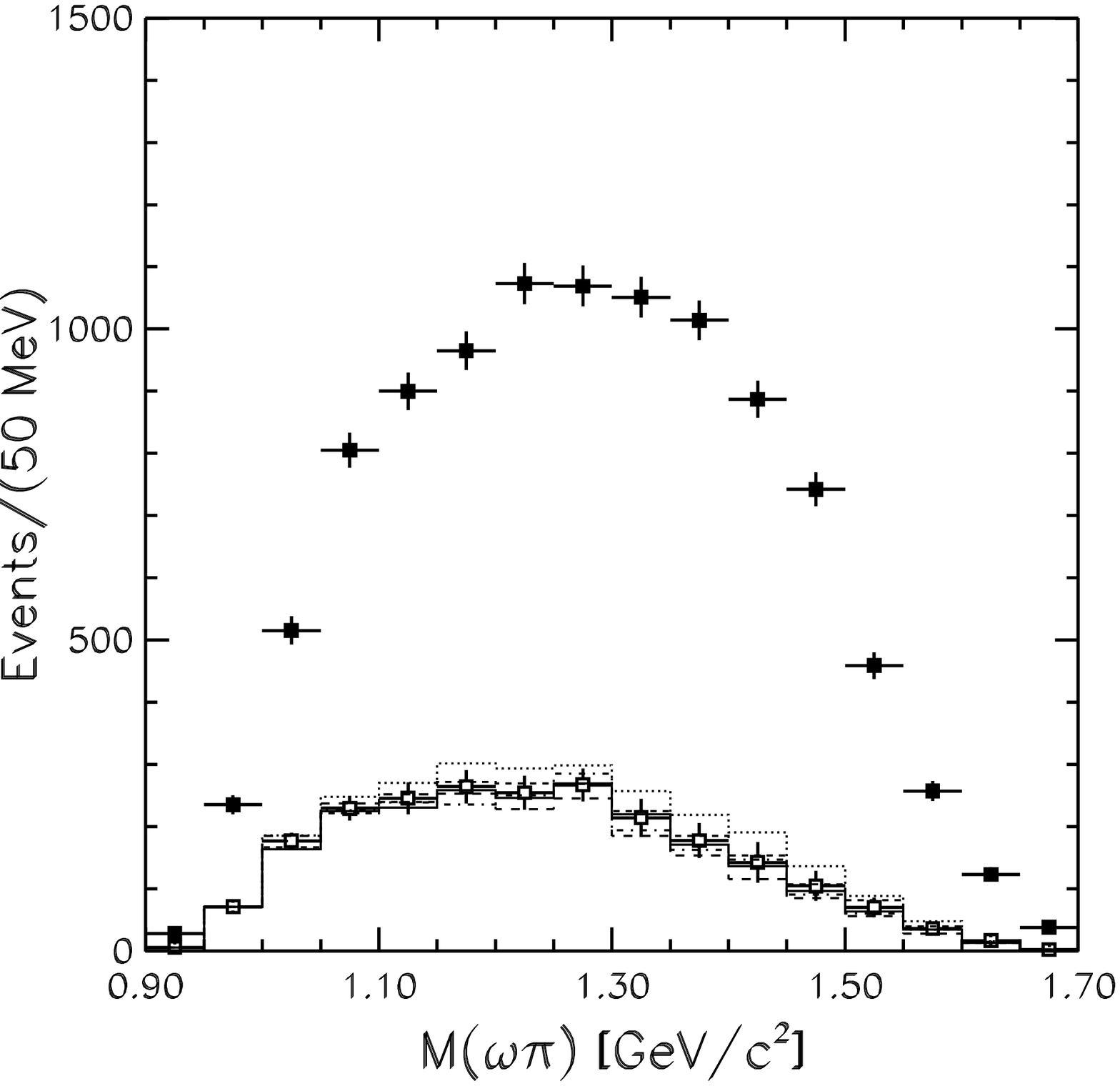}
{omegapi_spectral_bkg_variations}{
Distributions of the $\omega\pi$ effective mass for events from 
the
$\omega$ band (solid squares)
and the sidebands (open squares). The $\omega$ band distribution
includes combinatorial background as well.
Dotted and dashed lines indicate variations
in the shape of the combinatorial background 
due to variations in sideband definition.
The estimated average background shape, with errors, is shown as open 
squares.	
}

\simplex{htbp}{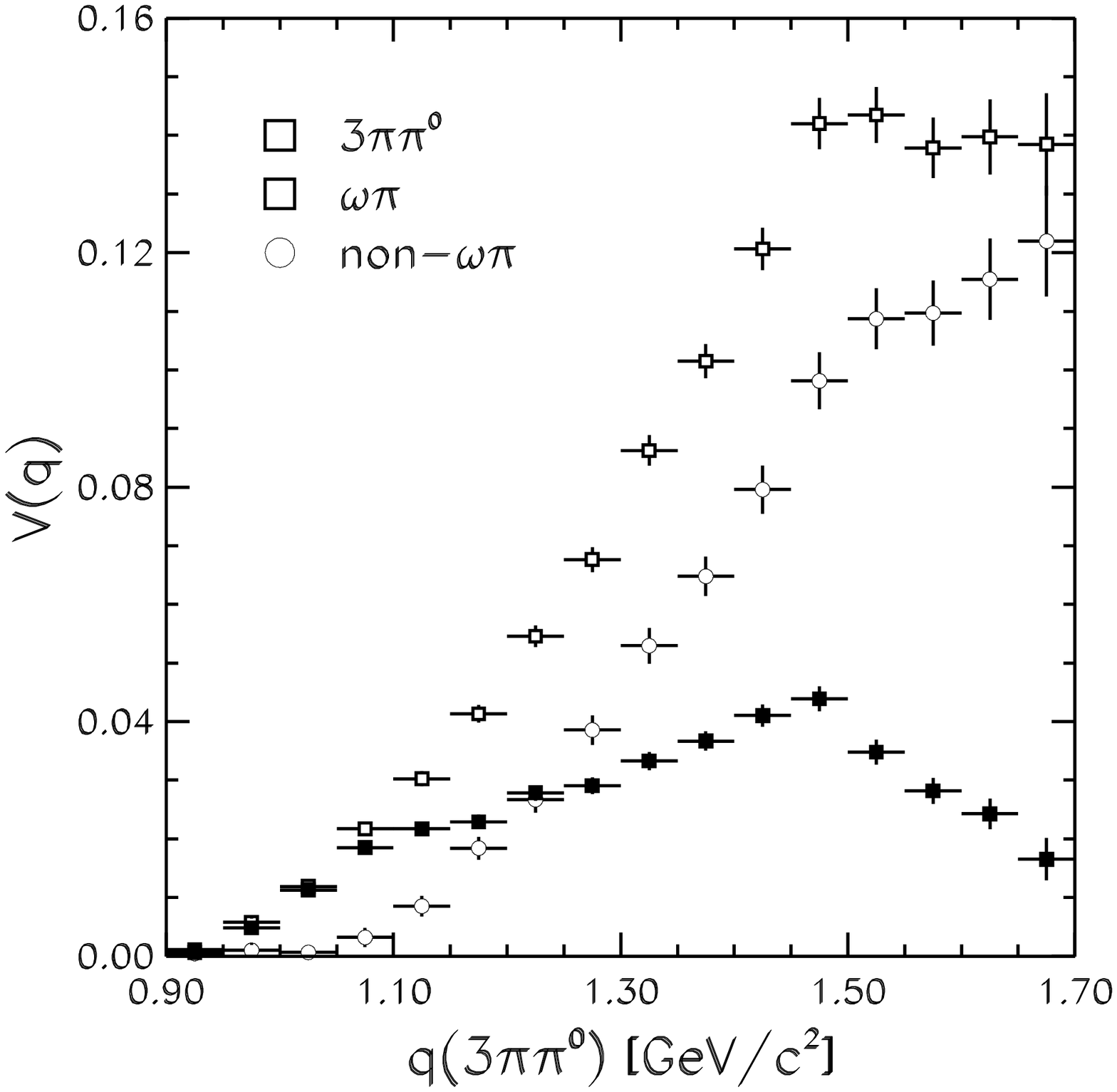}{omegapi_spectral}
{
 Spectral functions calculated for:  $3\pi\piz$ (open squares), 
$\omega\pi$ (solid squares) and non-$\omega\pi$ (open circles) 
components
}

\section{Model fits to the $\tau\to\omega\pi\nu_{\tau}$ spectral
function}
\label{sec:ompi_fit}

The spectral function of the $\omega\pi$ component 
can be expressed~\cite{tsai}
in terms of the weak form factor $F(q^2)$ as:
\begin{equation}
\label{eq:spec_ff}
V^{\omega\pi}(q) = \frac{1}{12\pi} \left(\frac{P_{\omega}(q)}{q}\right)^3 
\left|
\frac{g_{\rho\omega\pi}}{\gamma_{\rho}}
F(q) 
\right| ^2\ .
\end{equation}
Here, 
$g_{\rho\omega\pi}$ is the $\rho\to \omega\pi$ strong decay coupling,
$\gamma_{\rho}$
is a measure of the weak coupling
of the $\rho$ resonance to the weak charged current, and
\begin{equation}
\label{eq:pom}
P_{\omega}(q)=
\sqrt{
\frac{(q^2-(M_{\omega}-M_{\pi})^2)(q^2-(M_{\omega}+M_{\pi})^2 
)}{4q^2}}
\end{equation}
is the momentum of the $\omega$ in the $\omega\pi$ rest frame.

The choice of the form factor $F(q)$ is largely 
uncertain. In the region of $q^2\lsim M^2_{\tau}$ the form factor
$F(q)$ is expected to be dominated by low energy resonances. 
In this analysis we use a form for $F(q)$ that allows 
contributions from the $\rho$ resonance and its two
radial excitations $\rho'$ and $\rho''$:
\begin{equation}
F(q^2)=BW_{\rho}(q^2)+A_1 \cdot BW_{\rho '}(q^2)
	+A_2 \cdot BW_{\rho''}(q^2)\ .
\end{equation}
Here,
\begin{equation}
\label{eq:bw}
BW_{\rho}(q^2)=\frac{M^2_{\rho}}{M_{\rho}^2-q^2- i q \Gamma_{\rho}}
\end{equation}
is a relativistic Breit-Wigner amplitude normalized to unity at $q^2 = 0$,
and
\begin{equation}
A_i=
\frac{g_{\rho_{i}\omega\pi}}{g_{\rho\omega\pi}}
\frac{\gamma_{\rho}}{\gamma_{\rho_{i}}}
\end{equation}
are the ratios of the coupling constants for the different $\rho$ resonances.

To model    the Breit-Wigner shapes~(\ref{eq:bw}),
we first approximate $\Gamma_{\rho}$, $\Gamma_{\rho'}$ and
$\Gamma_{\rho''}$ by constants and do not include any momentum 
dependence.
The parameters of the $\rho(770)$ resonance have been throroughly studied 
and are known with great precision~\cite{PDG}. In contrast,
the parameters of the higher radial excitations 
$\rho'$ and $\rho''$ are poorly known and can be used as 
fit variables. Unfortunately, events above $1.6$ GeV/$c^2$
are suppressed by phase space and provide little sensitivity for 
the
heavy $\rho''$, the mass of which is expected to be around $1.7$ 
GeV/$c^2$.
In this study we only fit for the mass and width of $\rho'$ and 
keep widths
and masses of $\rho$ and $\rho''$ fixed at their central PDG
values~\cite{PDG}: 
\begin{center}
\begin{tabular}{rl}
$M_{\rho}=770$ MeV/c$^2$\ ,    & $\Gamma_{\rho}=151$ MeV/c$^2$\ ; 
\\
$M_{\rho''}=1700$ MeV/c$^2$\ , & $\Gamma_{\rho''}=235$ MeV/c$^2$\ 
.\\
\end{tabular}
\end{center}

The Conserved Vector Current (CVC) hypothesis predicts that the weak
form factor and the coupling constants are equal to the corresponding
electromagnetic quantities.
The weak coupling $W^\pm\to \rho^\pm\to \omega\pi^\pm$ in $\tau$ decays
is expected to be the same as the electromagnetic coupling
$\gamma^{*}\to \omega\pi^0$ in the reactions 
$e^+e^- \to \gamma^* (q_\gamma=\sqrt{s}) \to \omega\piz$
and $\omega\to\gamma(q_\gamma=0)\piz$.
All these couplings can be presented as $g_{\rho\omega\pi}/\gamma_\rho$,
where
\begin{equation}
\gamma_{\rho}=\sqrt{
\frac{4\pi\alpha^2M_{\rho}}{3\Gamma(\rho \to e^+ e^-)}}\ .
\end{equation}
Thus, one can use the Vector Dominance Model~\cite{dolinsky} to 
cross-check
the values of $A_1$ and $A_2$.
In the VDM, the coupling constants are energy independent, and therefore 
the set of fitted coupling constants should reproduce 
the experimentally measured value of:
\begin{equation}
\Gamma(\omega\to\piz\gamma)=\frac{1}{3}\alpha P_{\gamma}^3\cdot
\left|
\frac{g_{\rho\omega\pi}}{\gamma_\rho}
\left[
 1+A_1+A_2
\right]
\right|^2=(0.72\pm0.04)\ \mbox{MeV}/c^2,
\end{equation}
%
%
where $P_{\gamma}$ is the photon momentum in the $\omega$
center-of-mass system. 

To fit the data, we use the following spectral models:
\begin{itemize}
\item Model 1:
$F(q^2)\propto|BW_{\rho(770)}+A_1BW_{\rho'}+A_2BW_{\rho(1700)}|^2$\ .
\item Model 2:    $F(q^2)\propto|BW_{\rho(770)}+A_1BW_{\rho'}|^2$\ .
\item Model 3: 
$F(q^2)\propto|BW_{\rho(770)}+A_2BW_{\rho(1700)}|^2$\ .
\item Model 4:    $F(q^2)\propto|BW_{\rho(770)}|^2$\ .
\end{itemize}

The amplitudes $A_1$ and $A_2$ are assumed to be real.
We use $\chi^2$ minimization to fit various resonant models to
the $\omega\pi$ spectral function 
shown in Fig.~\ref{fig:omegapi_spectral_fits}.
To obtain the $\omega\pi$ spectral function 
shown in Fig.~\ref{fig:omegapi_spectral_fits}, we divide
the $\omega\pi$ spectral function 
shown in Fig.~\ref{fig:omegapi_spectral} by a value of the
branching fraction $\BR(\omega\to\pi^+\pi^-\piz)=0.888$ 
since the four pion final state does not include contributions
from any other decays of the $\omega$.
We verified that the fit results are independent of the 
chosen initial values by repeating the minimization 
procedure with different sets of initial values. 

The following parameters of the $\rho'$ are obtained from fits to the 
data with Models 1 and 2:
\begin{center}
\begin{tabular}{ccc}
 Model 1: & $M_{\rho'}=(1.52\pm0.01)$ GeV$/c^2$, &
 $\Gamma_{\rho'}=(0.38\pm0.04)$ GeV$/c^2$, \\
 Model 2: & $M_{\rho'}=(1.53\pm0.01)$ GeV$/c^2$, &
 $\Gamma_{\rho'}=(0.43\pm0.03)$ GeV$/c^2$. \\
\end{tabular}
\end{center} 
Using these results, we calculate weighted averages of the $\rho'$ 
parameters: $M_{\rho'}=(1.523\pm0.010)$ GeV$/c^2$,
$\Gamma_{\rho'}=(0.400\pm0.035)$ GeV$/c^2$. 

Models 3 and 4 do not provide an acceptable description of the data. 
Fits to our data corresponding to various combinations of the $\rho$
resonances are displayed in Fig.~\ref{fig:omegapi_spectral_fits},
and the results are summarized in Table~\ref{tab:fits}. 
We conclude that the presence of a $\rho'$ contribution is 
necessary to achieve an acceptable description of the data. 
Model 2 is favored, but Model 1 cannot be excluded.

\simplex{htbp}{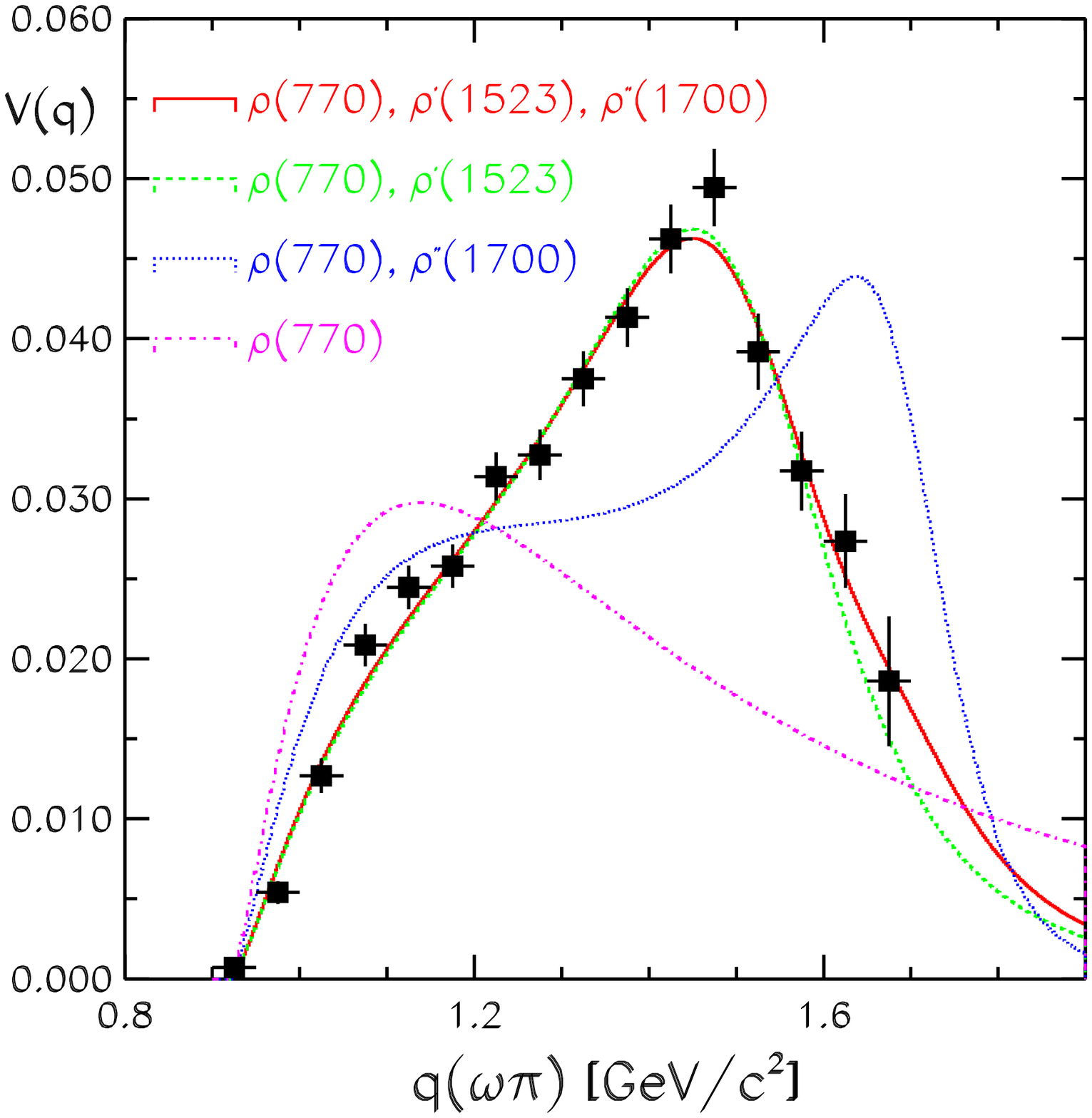}{omegapi_spectral_fits}
{Fits of the $\omega\pi$ spectral functions with various models.}

\begin{table}[bthp]
\caption{}{\label{tab:fits}
Fit results for various combinations of the $\rho, \rho', \rho''$ 
resonances
to the $\tau\to\omega\pi\nu_\tau$ spectral function.
Errors are statistical only and are strongly correlated.
Errors on the total width $\Gamma(\omega\to\piz\gamma)$
are obtained by propagating the entire covariance matrix.} 
     
\def\1#1#2#3{\multicolumn{#1}{#2}{#3}}
\renewcommand{\arraystretch}{0.8}
\begin{center}
\begin{tabular}{|l|r|r|r|r|r|}\hline
Model \# 
& g$_{\rho\omega\pi}$ & A$_1$ & A$_2$ & $\chi^2/d.o.f.$ 
& $\Gamma(\omega\to\piz\gamma)$ \\
     & $(\mbox{MeV}^2/c)^{-1/2}$ & & & & MeV \\ \hline
1
   & $16.4\pm0.6$ & $-0.20\pm0.04$ & $-0.017\pm0.016$
& 16.1/11 &  $0.87\pm0.15$   \\
\hline
2                    
   & $16.1\pm0.6$ & $-0.24\pm0.02$ & 
& 16.6/12 &  $0.78\pm0.10$   \\
\hline 
3 
   & $22.7\pm0.2$ &  & $-0.067\pm0.017$
& $342/14$ & $2.3\pm0.05$   \\
\hline
4                   
   & $27.2\pm0.2$ & & 
& $940/15$ & $3.9\pm0.03$   \\
\hline
\1{5}{|c|}{$\Gamma(\omega)\BR(\omega\to\piz\gamma)$} &
$0.72\pm0.04$ \\
\hline
\end{tabular}
\end{center}
\end{table}

\section{Inclusion of mass-dependent widths into the fit of the
$\tau\to\omega\pi\nu_\tau$ spectral function}

We now perform a fit to the
$\tau\to\omega\pi\nu_\tau$ spectrum assuming a mass-dependent width
for the three $\rho$ resonances. For this purpose, we use the results
of Section~\ref{sec:results}. We choose
the model that includes contributions from $\omega\pi$,
$\rho\pi\pi$, and non-resonant $3\pi\piz$ channels and use
the extracted values of $R_{\rho^0\pi^-\piz}$, $R_{\rho^-\pi^-\pi^+}$, 
$R_{\rho^+\pi^-\pi^-}$ 
and $R_{\omega\pi}$ to perform a numerical integration in
momentum space in order to obtain a spectral decomposition of
the $\tau\to 3\pi\piz\nu_\tau$ channel.
The results of the integration are shown
in Table~\ref{tab:ompi_and_rpp}. Here and below, we use the value of
the mass of the $\rho'$ extracted from the fits to the $\omega\pi$
spectral function as described in the previous Section.

\begin{table}[bthp]
\caption{}{\label{tab:ompi_and_rpp}
Contributions to the spectral functions of the 
$\omega\pi$ and $\rho\pi\pi$ subchannels of $\tau\to 3\pi\piz\nu_\tau$.} 
\begin{center}
\begin{tabular}{|c|c|c|} \hline
			& $\omega\pi$	& $\rho\pi\pi$	\\ \hline
non-resonant		& 7\%		& 76\%		\\ \hline
$\rho(770)$		& 36\%		& 0.2\%		\\ \hline
$\rho'(1520)$		& 12\%		& 5\%		\\ \hline
constructive interference & 47\%	& 20\%		\\ \hline	
\end{tabular}
\end{center}
\end{table}

With a better understanding of the $\omega\pi$ and
$\rho\pi\pi$ spectral functions, we perform a fit to the
$\tau\to\omega\pi\nu_\tau$ spectrum assuming mass-dependent widths
for the three $\rho$ resonances. Instead of constant-width
Breit-Wigner shapes~(\ref{eq:bw}), we use mass-dependent total
widths $\Gamma_{i}(q^2);\ 
i=\rho,\rho',\rho''$. To simulate the mass-dependent
widths, we follow the approach adopted in Ref.~\cite{kuhn}.

Since $\rho\to\pi\pi$ is the dominant $\rho$ decay mode, we take 
into
account only this channel for the calculation of $\Gamma_{\rho}(q^2)$:
\begin{equation}
\Gamma_{\rho}(q^2) = \Gamma_{\rho 0} \frac{M_{\rho}}{q} 
\left( \frac{p_{\pi}(q^2)}{p_{\pi}(M_{\rho}^2)} \right)^3\ ,
\end{equation}
where $\Gamma_{\rho 0}=151$~MeV and $M_{\rho}=770$~MeV are the 
central
width and mass of the $\rho$ resonance fixed at their PDG 
values~\cite{PDG}, 
$q$ is the
hadronic four-momentum, $p_{\pi}(q^2) = \sqrt{q^2/4 - M_{\pi}^2}$ is the
pion momentum in the $\rho$ rest frame, and $p_{\pi}(M_{\rho}^2)$
is the same momentum evaluated at the pole mass of the $\rho$ 
resonance.

For the total width of the $\rho'$ resonance, we take into
account only 
contributions from $\rho'\to\pi\pi$, $\rho'\to\omega\pi$ and
$\rho'\to\rho\pi\pi$ modes: 
\begin{eqnarray}
\label{eq:gamma_rhop}
\Gamma_{\rho'}(q^2) = \Gamma_{\rho' 0}  
\left[ 
\BR(\rho'\to\pi\pi) \frac{M_{\rho'}}{q}
\left( \frac{p_{\pi}(q^2)}{p_{\pi}(M_{\rho'}^2)} \right)^3
+
\BR(\rho'\to\omega\pi) \frac{M_{\rho'}}{q}
\left( \frac{p_{\omega}(q^2)}{p_{\omega}(M_{\rho'}^2)} \right)^3
\right. \nonumber \\ + \left.
\BR(\rho'\to\rho\pi\pi) 
\frac{F_{\rho\pi\pi}(q^2)}{F_{\rho\pi\pi}(M_{\rho'}^2)}
\right]\ ,
\end{eqnarray}
where $\Gamma_{\rho' 0}$ and $M_{\rho'}$ are the central
width and mass of the $\rho'$ resonance that are allowed to float 
in
the fit, $\BR(\rho'\to\pi\pi)$, $\BR(\rho'\to\omega\pi)$ and
$\BR(\rho'\to\rho\pi\pi)$ are 
relative branching fractions normalized with 
their sum is equal to one, and
$p_{\omega}(q^2)$ is the momentum of the $\omega$ in the 
$\omega\pi$
rest frame~(\ref{eq:pom}).
To extract the shape of $F_{\rho\pi\pi}(q^2)$, we generate a
$\tau\to \rho\pi\pi\nu_\tau$ sample with overall phase-space and 
$\omega\pi$
contributions eliminated. The form factor is defined as:
\begin{equation}
F_{\rho\pi\pi}(q^2) = \frac{1}{ (M_{\tau}^2+2q^2) 
(M_{\tau}^2-q^2)^2 }
\frac{1}{q} \frac{ dN }{dq}\ ,
\end{equation}
where $dN/dq$ is the number of Monte-Carlo events in a given bin of
the invariant mass $q$. 
To obtain the shape parameters, we fit the form factor to a 
second order polynomial.  

The choice of the relative branching fractions of $\rho'$ is 
largely uncertain~\cite{PDG}. 
The ratio $\BR(\rho'\to\rho\pi\pi) /
\BR(\rho'\to\omega\pi)$ can be extracted from our data. 
The relative contributions of the
$\rho'$ resonance into the $\omega\pi$ and
$\rho\pi\pi$ spectral functions, shown in
Table~\ref{tab:ompi_and_rpp}, are 12\% and 5\% respectively. Hence,
the ratio of 
the branching fractions is estimated as:
$$
\frac{\BR(\rho'\to\rho\pi\pi)}{\BR(\rho'\to\omega\pi)} = 
\frac{0.05}{0.12}
\frac{\BR(\tau\to\rho\pi\pi\nu_\tau)}{\BR(\tau\to\omega\pi\nu_\tau)
}\ ,
$$
where the ratio
${\BR(\tau\to\rho\pi\pi\nu_\tau)}/{\BR(\tau\to\omega\pi\nu_\tau)}$ 
is given by the results of our multidimensional fit described in 
Section~\ref{sec:results}.
We obtain 
$\BR(\rho'\to\rho\pi\pi) / \BR(\rho'\to\omega\pi) = 0.61$ 
and vary the value of
$\BR(\rho'\to\pi\pi)$ in the fit to make sure that the outcome 
depends weakly on the input value. For the final result shown 
in Fig.~\ref{fig:2rhos_mdw}, we use the constraint~\cite{PDG}
$\BR(\rho'\to\pi\pi) / \BR(\rho'\to\omega\pi) \sim 0.32$.

To simulate the mass-dependent width of the $\rho''$, 
we account only for the $\rho''\to\rho\pi\pi$ decay, 
since PDG~\cite{PDG} suggests that this is the dominant mode.
The mass-dependent width of the $\rho''$ is calculated similarly to
the $\rho'\to\rho\pi\pi$ contribution: 
\begin{equation}
\Gamma_{\rho''}(q^2) = \Gamma_{\rho'' 0} 
\frac{F_{\rho\pi\pi}(q^2)}{F_{\rho\pi\pi}(M_{\rho''}^2)}\ ,
\end{equation}
where $\Gamma_{\rho'' 0}=1700$~MeV and $M_{\rho''}=235$~MeV are the 
central
width and mass of the $\rho''$ resonance fixed at their PDG
values~\cite{PDG}, and $F_{\rho\pi\pi}(q^2)$ has the same
shape as for $\Gamma_{\rho'}(q^2)$.

Assuming this model, we fit the $\tau\to\omega\pi\nu_\tau$ spectral
function to the data. We make two fits: 1) the contribution
of the $\rho''$ resonance is set to zero, and 2) the contribution
of the $\rho''$ is allowed to float in the fit. In both situations the
extracted mass of the $\rho'$ is close to that of the $\rho''$. Hence, we
cannot distinguish between the two resonances, and we
assume that the shape of the spectral function is affected by the
$\rho'$ resonance only. The extracted mass and width of the $\rho'$ are
$1.67\pm0.03$~GeV and $0.8\pm0.1$~GeV, respectively. The fit is
shown in Fig.~\ref{fig:2rhos_mdw}. 

\simplex{htbp}{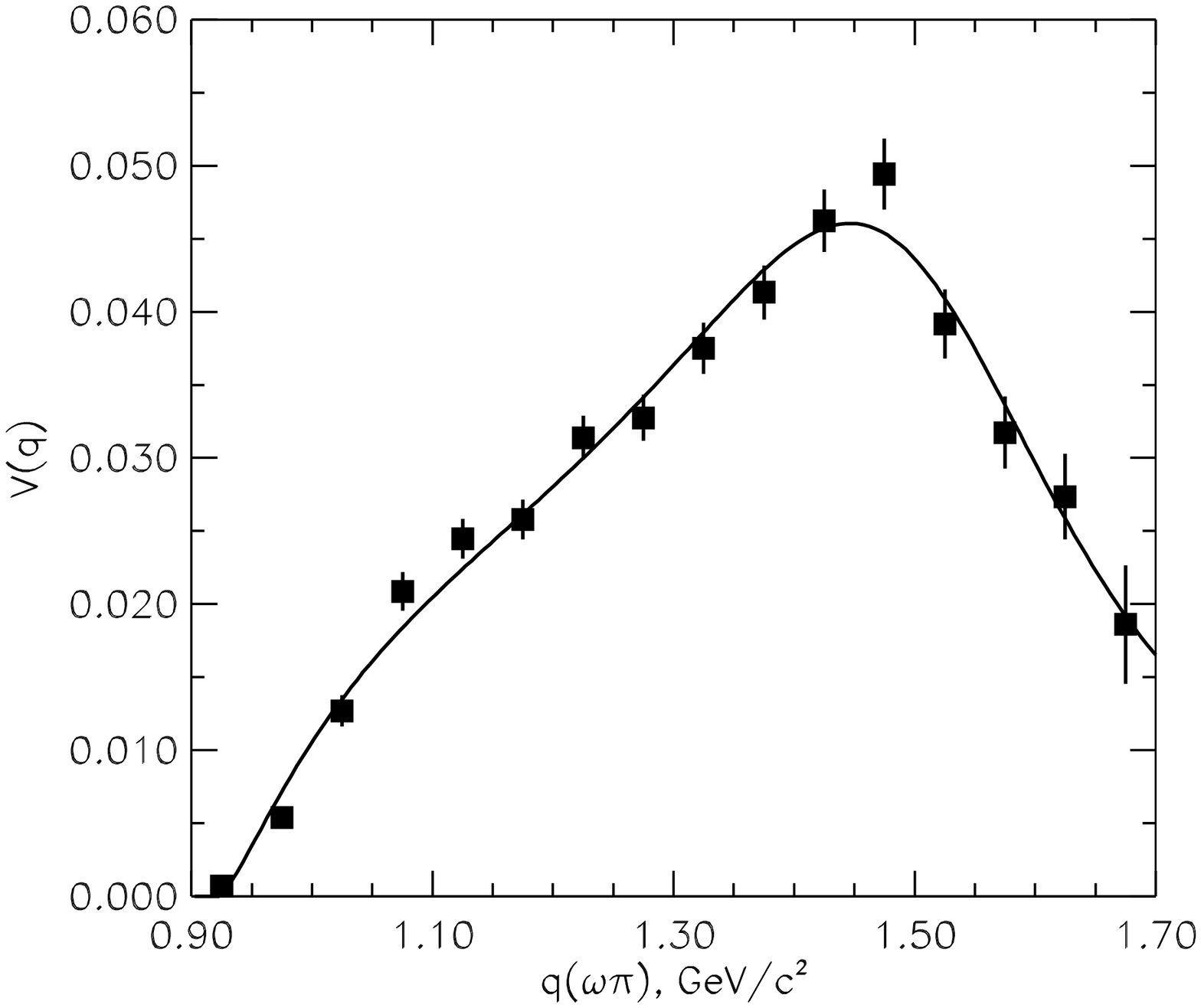}{2rhos_mdw}
{ Fit of the $\omega\pi$ spectral function with mass-dependent 
width.}

We also fit for the $\rho'$ and $\rho''$ contributions fixing the
masses and widths of $\rho'$ and $\rho''$ to their PDG values. The
obtained fit does not reproduce the shape of the measured spectral 
function,
which implies that the given
combination of $\rho'$ and $\rho''$ parameters does not provide an
adequate description of the data.

Finally, we attempt to fit for the relative phases
of $A_1$ and $A_2$, i.e., of
$\rho'$ and $\rho''$
contributions, respectively. Our study
shows that, varying the phases within $\pm\pi/4$, we do not 
introduce
any significant changes in the shape of the spectral function. 
Hence, we can argue neither in
favor of non-zero phases, nor in favor of phaseless models, and we
set the phases to zero for simplicity.

We also fit the $\tau\to\omega\pi\nu_\tau$ spectral function,
assuming another parameterization of the total width of the $\rho'$
resonance. Now the $\rho'\to\rho\pi\pi$ contribution is replaced
by that of $\rho'\to a_1\pi$ because Section~\ref{sec:results} suggests that
the four pion spectrum is dominated by the $\omega\pi$ and $a_1\pi$ 
channels. The last term in Eqn.~(\ref{eq:gamma_rhop}) is therefore 
replaced with
$\BR(\rho'\to a_1\pi) (M_{\rho'}/q) (p_{a_1}(q^2)/p_{a_1}(M_{\rho'}^2))$,
where $\BR(\rho'\to a_1\pi)$ is estimated using the results of 
Section~\ref{sec:results}. The extracted values of the mass and width 
of the $\rho'$ are similar to those quoted in this Section, i.e., 
significantly larger than the values obtained with the mass-independent
fit.

Because fits assuming mass-dependent widths rely heavily on the
choice of the model for the width parameterization, we use the mass
and width of 
$\rho'$ obtained in the fits with the mass-independent width
(Section~\ref{sec:ompi_fit}).
The analysis
described in this Section is presented mostly as a cross-check.

\section{Search for second class currents in $\tau\to\omega\pi\nu_{\tau}$ 
decay}

The decay $\tau\to \omega\pi\nu_\tau$
is expected to 
proceed through the hadronic vector current mediated by 
$\rho$, $\rho'$, $\rho''$ and higher excitations.
If, however, G parity
conservation is broken due to second class currents~\cite{ref:secondclass},
the decay can proceed
through a hadronic axial-vector
current mediated, e.g., by the $b_1(1235$) resonance. The 
difference in
spin-parity  
assignments for each of these states is reflected in 
different polarizations of the $\omega$ spin and hence in
different expected angular distributions
of $\cos\chi$. The angle $\chi$ is defined as the angle between the 
normal 
to the $\omega$ decay plane and the direction of the fourth pion
measured in the $\omega$ rest frame, and $L$ is the orbital angular
momentum of the $\omega\pi$ system.
The expected forms~\cite{chung} of the $\cos\chi$ distribution are
listed in Table~\ref{tab:shapes}.

\begin{table}[ht]
\caption{}{\label{tab:shapes} Expected shapes for different
spin-parity assignments in the decay $\tau\to\omega\pi\nu_{\tau}$.}
\begin{center}
\vspace{0.15cm}
\begin{tabular}{||c|c|c||}
\hline
$\hskip +1cm J^P \hskip +1cm$ & $\hskip +1cm L \hskip +1cm $ &
$\hskip +1cm F(\cos\chi) \hskip +1cm $ \\  \hline
$1^-$ & $1$ & $1-\cos^2\chi$ \\ \hline
$1^+$ & $0$ & $1$ \\ \hline
$1^+ $ & $ 2$  &  $ 1+3\cos^2\chi$ \\ \hline
$0^-$  & $ 1$  &  $ \cos^2\chi$  \\ \hline
\hline
\end{tabular}
\end{center}
\end{table}

We form the distribution of $\cos\chi$ for the
$\tau\to\omega\pi\nu_\tau$ events in our data.
To eliminate combinatorial and non-$\omega$ background, we perform
a sideband subtraction for every bin, in the same way as described in
Section~\ref{sec:spectr}. 
Using Monte Carlo estimates, we subtract small $\omega$ 
contributions 
from $\tau$ and $q\bar{q}$ 
background events, correct for efficiency,
and fit the resulting distribution, shown in Fig.~\ref{fig:cosn1},
with the following function: 
\begin{equation}
F=(1-\epsilon)\cdot F^1+\epsilon \cdot F^2\ .
\end{equation} 
Here, $F^1(\cos\chi)=1-\cos^2\chi$ is the 
shape of the dominant 
vector contribution and  
$F^2(\cos\chi)=1$ represents the shape which gives 
the most conservative estimate of the 
non-standard contribution
to the  $\tau\to\omega\pi\nu_{\tau}$ decay. 
The value of $\epsilon$ obtained from the fit is 
consistent with zero 
within errors: $\epsilon=(0.08\pm2.00)\times10^{-2}$. After 
integration, it 
can be translated into an upper limit on the ratio
$N^{\omega\pi}\mbox{(non-vector current)}
/ N^{\omega\pi}\mbox{(vector current)} < 5.4\%$
at 90\% CL and 
$< 6.4\%$ at 95\% CL, where
$N^{\omega\pi}$ represents the number of events generated through 
the
corresponding current.
An analogous study~\cite{aleph} by ALEPH yielded
$N^{\omega\pi}\mbox{(non-vector current)} 
/ N^{\omega\pi}\mbox{(vector current)}<8.6$\% at 95\% CL.  
The solid line in Fig.~\ref{fig:cosn1} illustrates the
agreement of the $(1-\cos^2\chi)$ description and the data.


\simplex{htbp}{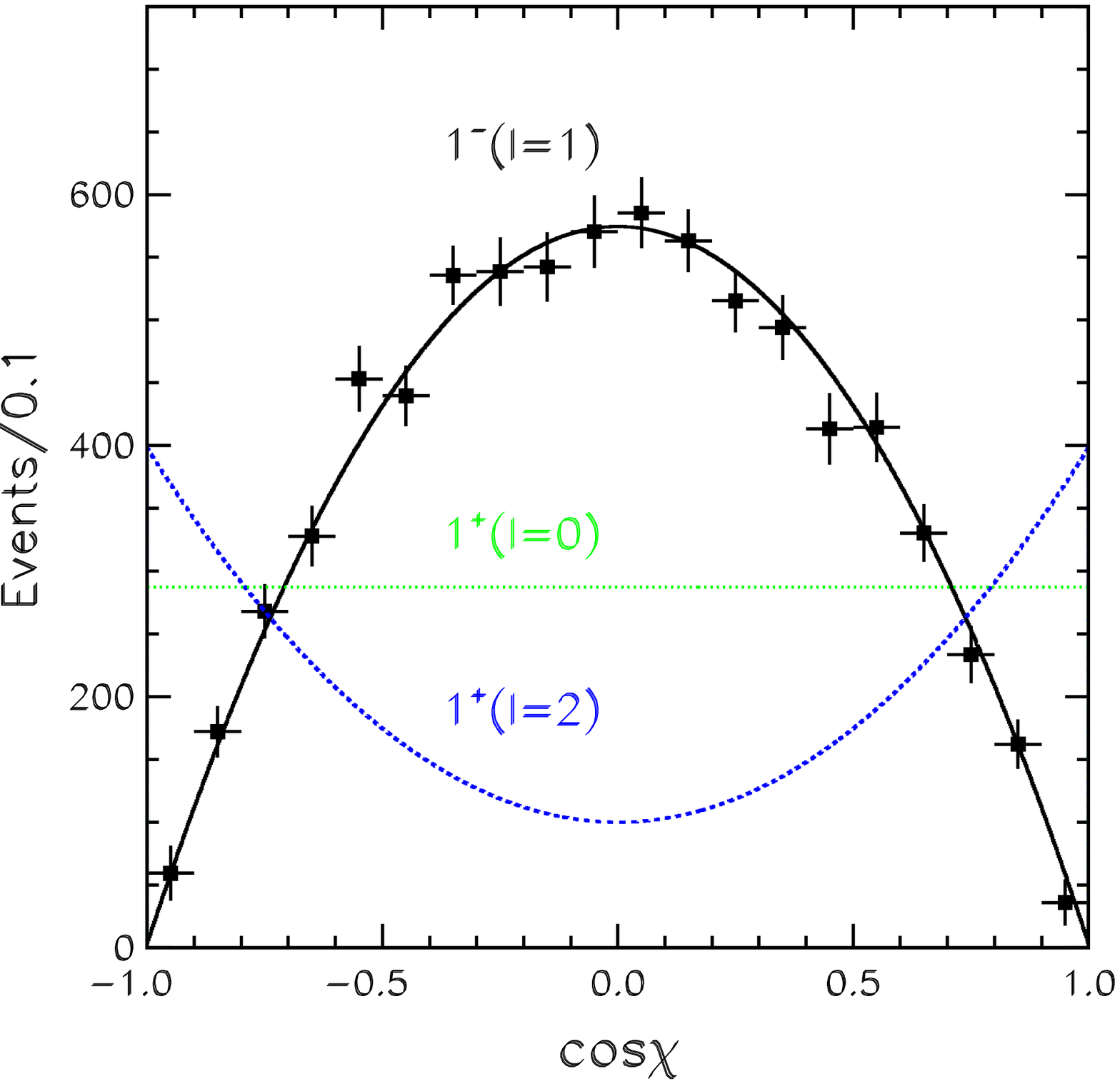}{cosn1}
{
The distribution of $\cos\chi$ for $\tau\to\omega\pi\nu_{\tau}$ 
events
after
background subtraction and efficiency correction is shown
as filled squares with error bars.
The solid line 
represents a
fit to the vector current model $J^P_L=1^{-}_{1}$, 
the dashed line shows
the expected 
shape
for $J^P_L=1^{+}_{2}$, and the dotted
line shows the $J^P_L=1^{+}_{0}$ model.
}

\section{Full unbinned likelihood fit to the $\tau\to 
3\pi\piz\nu_{\tau}$ resonant structure}

A general method of obtaining a resonant decomposition of a hadronic 
final state is a partial wave analysis of the data via
an unbinned maximum likelihood fit. 
For any data set with limited statistics, the analysis is model-dependent 
since there is a necessary choice of amplitudes to be considered.
In this study, we consider four models with contributions from
$\omega\pi$, $a_1\pi$, $\sigma\rho$, $f_0\rho$, $\rho\pi\pi$, and non-resonant 
$3\pi\piz$ channels.

Results depend on
the model chosen for the hadronic current parameterization. 
A general description and some details of the formalism are 
given below. The amplitude description is given in the Appendix.

\subsection{Definition of the fit function}

The differential width of the $\tau$ decay into the $3\pi\piz$ 
final state is 
described by a factorizable sequence of decays: $\tau\to h + \nu_\tau$ 
followed by $h\to 4\pi$, and can be expressed as:

\begin{equation}
d\Gamma(\tau\to3\pi\piz\nu_{\tau})=\frac{1}{2M_{\tau}}|{\mathcal{M}}|^2
d\mbox{LIPS}(\tau\to\nu_{\tau}h)d\mbox{LIPS}(h\to 3\pi\piz)\ ,
\end{equation}
where 
\begin{equation}
|{\mathcal{M}}|^2=\frac{G^2_F}{2}V^2_{ud}{ L^{\mu\nu}H_{\mu\nu}}
\end{equation}
is the square of the corresponding matrix element, $d\mbox{LIPS}$ 
is an
element of the phase space volume, and $L^{\mu\nu}$ and $H_{\mu\nu}$
are lepton and hadronic tensors, respectively. 

The lepton tensor $L^{\mu\nu}$ cannot be calculated without 
information on the neutrino direction. To resolve this problem, we use 
a general form of the matrix element proposed 
by Mirkes and K\"uhn~\cite{mirkes}
which is explicitly averaged over the neutrino direction. 
In this approach the 
averaged matrix element is expressed as a sum of 16 terms:
\begin{equation}
\overline{L^{\mu\nu}H_{\mu\nu}}=2(M^2_{\tau}-q^2)\sum_{i=1}^{16}
\overline{L_i}W_i\ , 
\end{equation}
where the weights $W_i$ can be calculated from components of the complex 
hadronic current $J_{\mu}$,
and $\overline{L_i}$ are kinematic parameters which depend only on the
four-momenta of the four pions in the $3\pi\piz$ final state. 

To calculate $J_{\mu}$, we 
use the {\tt KORALB} Monte Carlo program with  
the {\tt TAUOLA} decay package~\cite{KORALB}. 
The reconstructed four-momenta
of the $2\pi^-\pi^+\piz$ system are boosted into the rest frame of 
the hadronic system which is 
denoted by the superscript ``$cm$''. They are then projected onto 
the following coordinate 
system: the $x$-axis is along the direction of
$(\vec{P}^{cm}_{\pi^+}+\vec{P}^{cm}_{\piz})$, 
the $z$-axis is perpendicular to the plane defined by the vectors
$\vec{P}^{cm}_{\pi_1^-}$ and $\vec{P}^{cm}_{\pi_2^-}$, 
and the $y$-axis is perpendicular to the $x-z$ plane. We found that
the calculated weights 
reveal large unphysical fluctuations unless we require 
$\cos\psi>-0.97$, 
where $\psi$ is the angle between the direction of the laboratory 
and the $\tau$, as seen from the hadronic rest frame \cite{kuhn}.

To check whether the calculation of the 
$\overline{L^{\mu\nu}H_{\mu\nu}}$ 
expression is correct, we choose a parameterization of the hadronic current 
with form factors
and coupling constants that provide a good description of the data. 
We generate a large sample of $\tau\to 3\pi\piz\nu_{\tau}$ events 
according to the  phase space and for each event we calculate the 
corresponding averaged weight: 
$\overline{L^{\mu\nu}H_{\mu\nu}}$. As a cross check, 
we use the same parameterization of the
hadronic current to generate another sample of unit weight events using the 
standard full {\tt TAUOLA}
simulation. The full $\tau\to 3\pi\piz\nu_{\tau}$ simulation 
invokes explicit generation
of neutrino momenta and subsequent calculation of the precise matrix element 
$L^{\mu\nu}H_{\mu\nu}$. In the limit of large statistics, both 
methods should give the 
same results. The shapes of submass and 
total hadronic energy distributions 
generated with both methods  
agree, though the distributions obtained with 
the averaged matrix
element reveal a higher level of fluctuations. 
This is due to the fact that the 
averaging over the neutrino direction 
introduces an additional smearing which increases errors.


\subsection{Likelihood function} 
\label{sec:likelihood}

The matrix element averaged over the neutrino direction 
can be used as a probability
density function (p.d.f.) to fit the data of the $\tau\to 
3\pi\piz\nu_{\tau}$ decays.
The fitting technique is a straightforward 
extension of that used to fit three-body Dalitz plots. 

We take advantage of the following convenient notation~\cite{rhovsrho}.
The matrix element ${\mathcal{M}}(\vec{\alpha},\vec{\beta}|\vec{\theta})$
is described by a model parameter vector $\vec{\theta}$ which consists of 
resonant amplitudes and phases. The matrix element  
is a function of observable variables $\vec{\alpha}$ (e.g., 
hadronic four-momenta)  
and non-observable variables $\vec{\beta}$ (e.g., neutrino four-momentum). 
The matrix element has to be integrated over the non-observable variables: 
\begin{equation}
\label{eq:f_s}
f_s(\vec{\alpha}|\vec{\theta})=\int
|{\mathcal{M}}(\vec{\alpha},\vec{\beta}|\vec{\theta})|^2 d\vec{\beta}\ ,
\end{equation}
where $f_s(\vec{\alpha}|\vec{\theta})$ is the probability
to observe a final state with a parameter vector $\vec{\alpha}$,
given the model parameters $\vec{\theta}$.
To compare directly with the data distribution,
$f_s$ must be 
multiplied by the phase space function $\phi(\vec{\alpha})$ and by the 
detector acceptance $\eta(\vec{\alpha})$: \\
\begin{equation}
f_s^{exp}(\vec{\alpha}|\vec{\theta})=\eta(\vec{\alpha})\phi(\vec{\alpha}) 
f(\vec{\alpha}|\vec{\theta})\ .
\end{equation}
Since we are trying to determine parameters of the matrix element 
and have all observables fixed by data, we follow the usual Bayesian
practice and exchange the argument and the parameter
in the expression for $f_s^{exp}$,
thus interpreting it as a probability distribution for $\vec{\theta}$,
given the event observables $\vec{\alpha}$.

Since the parameters of the p.d.f are varied in the fit,
the normalization of the matrix element is no longer constant and
must be recalculated on each step of the fit:
\begin{equation}
F_s=\frac{f_s(\vec{\theta}|\vec{\alpha})}{N(\vec{\theta})}=
\frac{f_s(\vec{\theta}|\vec{\alpha})}{\int\eta(\vec{\alpha})\phi(\vec{\alpha}) 
f_s(\vec{\theta}|\vec{\alpha})d\vec{\alpha}}\ .
\end{equation}
Due to the presence of background, we must also include  
a corresponding background p.d.f. $F_b$, and define the total 
p.d.f. as: 
\begin{equation}
F=\frac{R_{S/B}F_s+F_b}{R_{S/B}+1}\ , 
\end{equation}
where  $R_{S/B}$ is a signal-to-background ratio. 
The likelihood function for selected events is defined as:
\begin{equation}
\label{eq:lkh_prod}
{\pounds(\vec{\theta})}=\prod_{events}F^{exp}=\prod_{events}
\frac{R_{S/B}F_s+F_b}{R_{S/B}+1} 
\eta(\vec{\alpha})\phi(\vec{\alpha})\ .
\end{equation}
To find the maximum of the likelihood, we minimize $-2Ln\pounds$, where
\begin{equation} 
Ln{\pounds(\vec{\theta})} =
 \sum_{j=1}^{N} Ln\frac{R_{S/B}F_s+F_b}{R_{S/B}+1} +
 \sum_{j=1}^{N} Ln(\eta(\vec{\alpha})\phi(\vec{\alpha}))\ . 
\end{equation}
The term $\sum Ln(\eta(\vec{\alpha})\phi(\vec{\alpha}))$ is constant 
for the purpose
of minimizing $F$ and is therefore  neglected. Thus, we never
need to explicitly evaluate 
$\eta(\vec{\alpha})$ and $\phi(\vec{\alpha})$: 
the amplitude normalization ${N(\vec{\theta})}$ obtained by Monte
Carlo techniques takes them into account automatically.    
When parameter terms can be factored out, the computation of the
normalization $N(\vec{\theta})$ can be speeded up: 
\begin{eqnarray} 
N(\vec{\theta})=\int\eta(\vec{\alpha})\phi(\vec{\alpha})
f(\vec{\theta}|\vec{\alpha})d\vec{\alpha}=
\int\eta(\vec{\alpha})\phi(\vec{\alpha})\left [\sum_j 
C_j(\vec{\theta})f_j(\vec{\alpha}) \right ] d\vec{\alpha}=
\nonumber \\
\sum_j C_j(\vec{\theta})\int\eta(\vec{\alpha})\phi(\vec{\alpha})
f_j(\vec{\alpha})d\vec{\alpha}\ ,
\end{eqnarray}
where the index $j$ runs over all possible combinations of the 
fit parameters $\vec{\theta}$.
During the fit, the integrals in the above 
expression are calculated only once and then 
multiplied by the coefficients $C_j$ which depend on the parameters
$\vec{\theta}$ only.
 
\subsection{Signal p.d.f.} 
\label{signal_pdf}

As shown in Section~\ref{sec:likelihood},
all constant factors and functions of observable variables $\vec{\alpha}$
that multiply the signal p.d.f.~(\ref{eq:f_s}) do not change the 
position of a maximum of the likelihood function~(\ref{eq:lkh_prod})
in the parameter space $\vec{\theta}$.
Neglecting such contributions,
the signal p.d.f~(\ref{eq:f_s}) can be expressed as:
\begin{equation}
f_s=\overline{L^{\mu\nu}H_{\mu\nu}}=\sum_{1}^{16}\overline{L_i} 
W_i\ .
\end{equation}
The hadronic current for the decay $\tau^-\to 
2\pi^-\pi^+\piz\nu_\tau$
is parameterized as follows:
\begin{equation}
\label{eq:hadcur}
J^{\mu}= \alpha_\omega f_\omega^\mu F_{\omega}(q) +   
         \sum_k \alpha_k f_k^\mu F_k(q)\ ,
\end{equation}
where $q$ is the total hadronic four-momentum. The
spectral functions (amplitude form factors) are defined as sums of the
corresponding Breit-Wigner amplitudes
\begin{equation}
\label{eq:fbw}
F_k(q)=\beta_k^0+\beta_k BW_{\rho}(q)+\beta'_k
BW_{\rho'}(q)+\beta''_k BW_{\rho''}(q)\ ;
\end{equation}
the coefficients
$f_k^{\mu}$ are functions of four-momenta describing the shapes of the
corresponding contributions, the coefficients $\alpha_k$ are complex
amplitudes, the coefficients $\beta_k$, $\beta'_k$ and 
$\beta''_k$ are complex amplitudes of the $k$th spectral function component,
and the sum in Eqn.~(\ref{eq:hadcur}) is taken over all non-$\omega$
contributions, as described in Section~\ref{sec:results} below.
For practical purposes, the amplitude of one of the $\alpha_k$ is set 
to a non-zero constant and the phase of one of the $\alpha_k$ is set to 
zero; the corresponding   
rescaling factors are then included into the  
other amplitudes, while the overall normalization of the signal
p.d.f. is set to unity.
The explicit form of the form factor $f_\omega^\mu$ is based
on Refs.~\cite{decker,KORALB,chiral},
which employs correct
$\omega$ 
helicity factors and proper symmetrization.

From the fit, we determine the amplitudes $\alpha_k$ for the decay
$\tau\to 2\pi^-\pi^+\piz\nu_\tau$. 
We integrate these amplitudes over a 
large sample of $\tau\to 3\pi\piz\nu_{\tau}$ 
events to translate them into relative branching fractions.

\subsection{Background p.d.f.} 

The total  background in the selected data sample is estimated
to be $4.9\%$. Because it is small, we do not calculate a p.d.f. for 
each individual component separately. Instead,  we use 
a simpler, empirical approach: 
we reweight the $3\pi\piz$ phase space to
create an effective matrix element describing the shape of the sum of
the $\tau$ and $q\bar{q}$ backgrounds 
that was determined using detailed Monte Carlo simulations.
We use a purely empirical form of this effective matrix element:
\begin{eqnarray}
F_b=(M_{\tau}^2-M(3\pi\piz)^2)\times(1+ \nonumber \\
     \frac{0.8}{(M(\piz\pi^-_1)^2-M_{\rho}^2)^2+\Gamma_\rho^2}+
     \frac{0.8}{(M(\piz\pi^-_2)^2-M_{\rho}^2)^2+\Gamma_\rho^2}+
\nonumber \\
\frac{0.011}{(M(\piz\pi^+\pi^-_1)^2-M_{\omega}^2)^2+\Gamma_\omega^2}+
\frac{0.011}{(M(\piz\pi^+\pi^-_2)^2-M_{\omega}^2)^2+\Gamma_\omega^2}+
\\
     \frac{0.005}{(M(\pi^-_1\pi^+)^2-M_{K_s}^2)^2+\Gamma_{K_s}^2}+
     \frac{0.005}{(M(\pi^-_2\pi^+)^2-M_{K_s}^2)^2+\Gamma_{K_s}^2})\ .
\nonumber
\end{eqnarray}
The choice of the parameterization is largely {\it ad hoc},
the coefficients are tuned to describe the background distributions 
in seven mass projections, as shown in Fig.~\ref{fig:back_param}.
These distributions 
are obtained from 
Monte Carlo simulations of $\tau$ backgrounds and
an appropriately scaled contribution from $e^+e^-\to q\bar{q}$ 
continuum events.
 
\simplex{htbp}{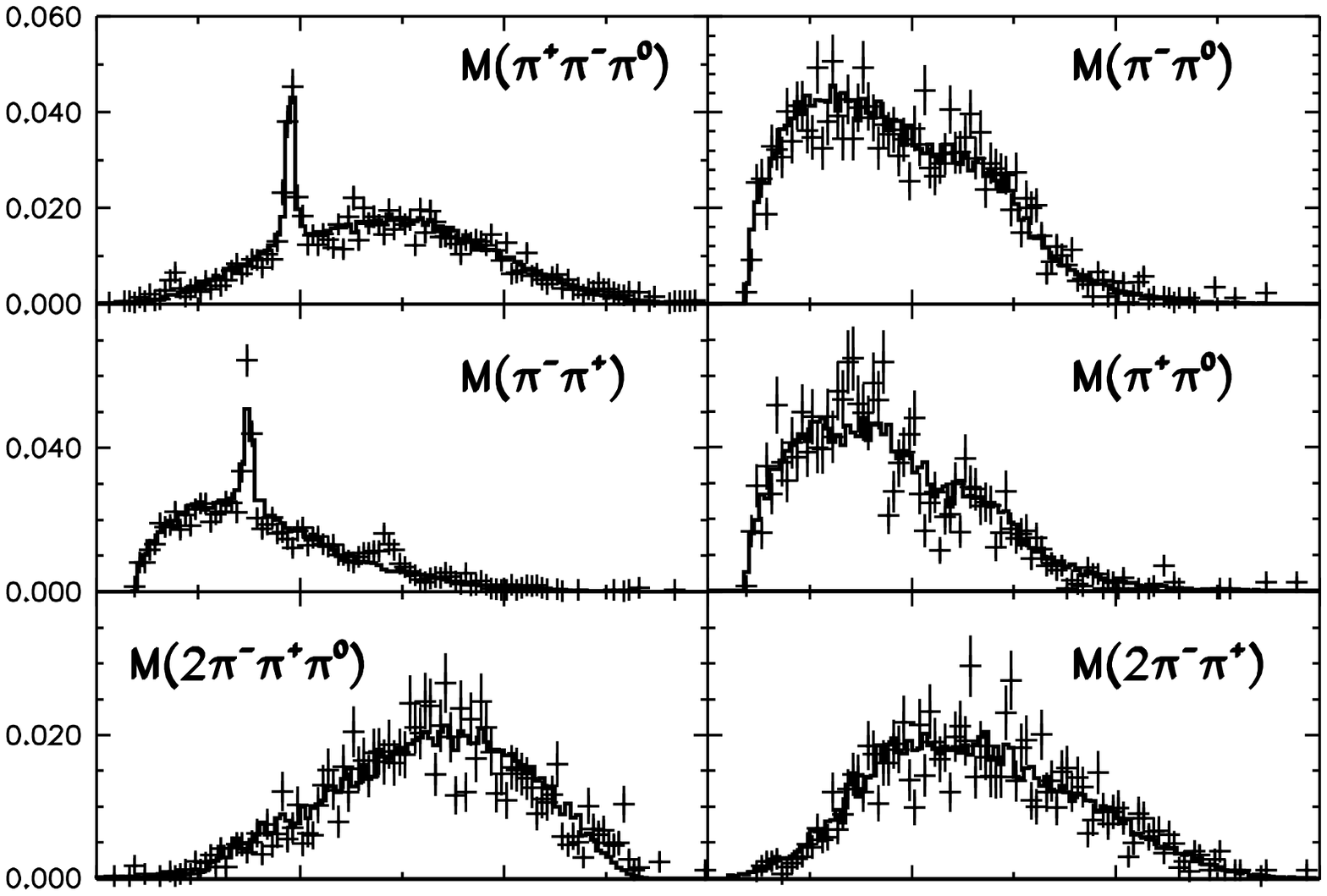}{back_param}
{Background parameterization plotted for different submass projections.
The predictions from Monte Carlo simulations are shown with crosses.
The background parameterization is 
shown as a histogram.}

\subsection{Fits to various models}
\label{sec:results} 

We use 4 models to parameterize the hadronic current. The models
account for the following contributions:
\begin{itemize}
 \item Model 1:  $\omega\pi$, $\rho\pi\pi$ and non-resonant $3\pi\piz$;
 \item Model 2:  $\omega\pi$ and $a_1\pi$;
 \item Model 3:  $\omega\pi$, $a_1\pi$, $\sigma\rho$ and $f_0(980)\rho$;
 \item Model 4:  $\omega\pi$, $a_1\pi$ and $\rho\pi\pi$.
\end{itemize}

Hadronic current expressions for each model are given in 
the Appendix. For the $\rho\pi\pi$ contribution, the spectral 
functions for
the $\rho^-\pi^+\pi^-$, $\rho^0\piz\pi^-$ and $\rho^+\pi^-\pi^-$ 
components
are assumed to be identical, and the corresponding form factors $f_i^\mu$
are parameterized in the same way as it is done in the {\tt TAUOLA} 
package~\cite{KORALB}.
We use the PDG~\cite{PDG} value for the mass 
of the $a_1$ resonance ($M_{a_1}=1230$~MeV) 
and adopt the prescription of Ref.~\cite{santamaria} to
use the higher end of the width interval
recommended by PDG ($\Gamma_{a_1}=600$~MeV).
For the mass and width of $\sigma$, i.e., $f_0(400-1200)$ in PDG's
notation, we assume values~\cite{tornq} of $M_\sigma=860$~MeV and 
$\Gamma_\sigma=880$~MeV.
We use PDG values for the mass and width of the $f_0(980)$ of 
$M_{f_0}=980$~MeV and $\Gamma_{f_0}=70$~MeV.
To obtain 
the relative contributions of the $\tau\to 3\pi\piz\nu_\tau$ decay 
components, the fit parameters are integrated over 300,000 of
$3\pi\piz$ phase space events and renormalized. 
Fit errors are propagated in accordance with this procedure.
We use the following definition:
$R_i = \BR_i / \BR(\tau^-\to 2\pi^-\pi^+\piz\nu_{\tau})$, where 
$\BR_i$ is a branching fraction for the corresponding channel, e.g., 
$R_{a_1\pi} = \BR(\tau^-\to (a_1\pi)^-\nu_\tau) / 
\BR(\tau^-\to 2\pi^-\pi^+\piz\nu_{\tau})$.
Statistical  errors on the relative fractions 
$R_i$ are obtained from the maximum likelihood fits assuming one 
standard deviation.

To simulate the form factors $BW_{\rho'}$, $BW_{a_1}$ and $BW_\sigma$,
we used both fits with mass-dependent and mass-independent widths.
The observed variation in the
amplitude values is small and has been included in the total
systematic error.
We fix the values of the mass and central width of 
$\rho'$ in the fit; if these
parameters are allowed to vary, the fit becomes unstable.

To estimate the goodness-of-fit for each model, we generate 40
Monte Carlo samples with the matrix element averaged over the neutrino
direction. Each sample is
generated with parameters obtained in the fit to the data and  
the number of events in each sample
is approximately equal to that observed in the data. 
The goodness-of-fit is estimated as a fraction of samples where
$-2Ln\pounds$ exceeds that in the data.

Integrated amplitudes with statistical and systematic errors, as 
well as the confidence levels of the fits for the four chosen models, are
listed in Table~\ref{tab:results}. The calculation of the systematic errors
is explained in the following section.
Projections of the fits on 
various mass combinations are shown in 
Figs.~\ref{fig:fits_1}-\ref{fig:fits_3}.

Models 3 and 4 provide the best description of the data. Both these
models are dominated by $\omega\pi$ and $a_1\pi$
with small additional contributions of the $\sigma\rho$, $f_0\rho$ or
non-resonant $\rho\pi\pi$ channels.
Submass projections for Models 3 and 4 are almost identical, and we show
only those for Model 3.

\begin{table}[bthp]
\caption{}{\label{tab:results}
Fit results for various models. $R_i$ is defined as:
$R_i = \BR_i / \BR(\tau^-\to 2\pi^-\pi^+\piz\nu_{\tau})$, e.g., 
$R_{a_1\pi} = \BR(\tau^-\to (a_1\pi)^-\nu_\tau) / 
\BR(\tau^-\to 2\pi^-\pi^+\piz\nu_{\tau})$.}
\begin{center}
\begin{tabular}{|c|c|c|c|} \hline
Model 		& Integrated amplitudes & Sum of amplitudes & 
Goodness-of-fit \\ \hline
& $R_{\rho^0\pi^-\piz}=0.11\pm0.01\pm0.02$ &  & \\
& $R_{\rho^-\pi^-\pi^+}=0.19\pm0.02\pm0.04$ &  & \\
Model 1
& $R_{\rho^+\pi^-\pi^-}=0.23\pm0.02\pm0.04$ & $0.93\pm 0.04\pm 0.08$ 
& $15\%$ \\
& $R_{\omega\pi}=0.40\pm0.02\pm0.05$ &  & 
\\
& $R_{3\pi\piz}<0.06$ at 95\% CL &  & 
\\ \hline
Model 2 
& $R_{\omega\pi}=0.38\pm0.02\pm0.02$ & $0.81\pm 0.03\pm 0.03$ & $<5\%$ \\
& $R_{a_1\pi}=0.43\pm0.02\pm0.02$ &  &  \\ \hline
& $R_{\omega\pi}=0.38\pm0.02\pm0.01$ &  & \\
Model 3
& $R_{a_1\pi}=0.49\pm0.02\pm0.02$ & $0.89\pm 0.03\pm 0.03$ & $20\%$ \\
& $R_{\sigma\rho}=0.01\pm0.02\pm0.01$ &  & \\
& $R_{f_0\rho}=0.01\pm0.01\pm0.01$    &  & \\ \hline
& $R_{\omega\pi}=0.39\pm0.02\pm0.01$ & & \\
& $R_{a_1\pi}=0.50\pm0.03\pm0.01$ & &  \\
Model 4
& $R_{\rho^0\pi^-\piz}=0.01\pm0.01\pm0.01$ & $0.93\pm 0.04\pm 0.02$ & $20\%$ \\
& $R_{\rho^-\pi^-\pi^+}=0.02\pm0.03\pm0.01$ &  & \\
& $R_{\rho^+\pi^-\pi^-}=0.01\pm0.01\pm0.01$ &  & \\ \hline
\end{tabular}
\end{center}
\end{table}

The presence of a non-resonant $3\pi\pi^0$ contribution does 
not improve the 
goodness-of-fit in the 
$\omega\pi$-$\rho\pi\pi$-non-resonant-$3\pi\piz$ model, and
we set an upper limit on the non-resonant
contribution shown in Table~\ref{tab:results}. We also ignore the
non-resonant $3\pi\piz$ component in the two other models.

\simplex{htbp}{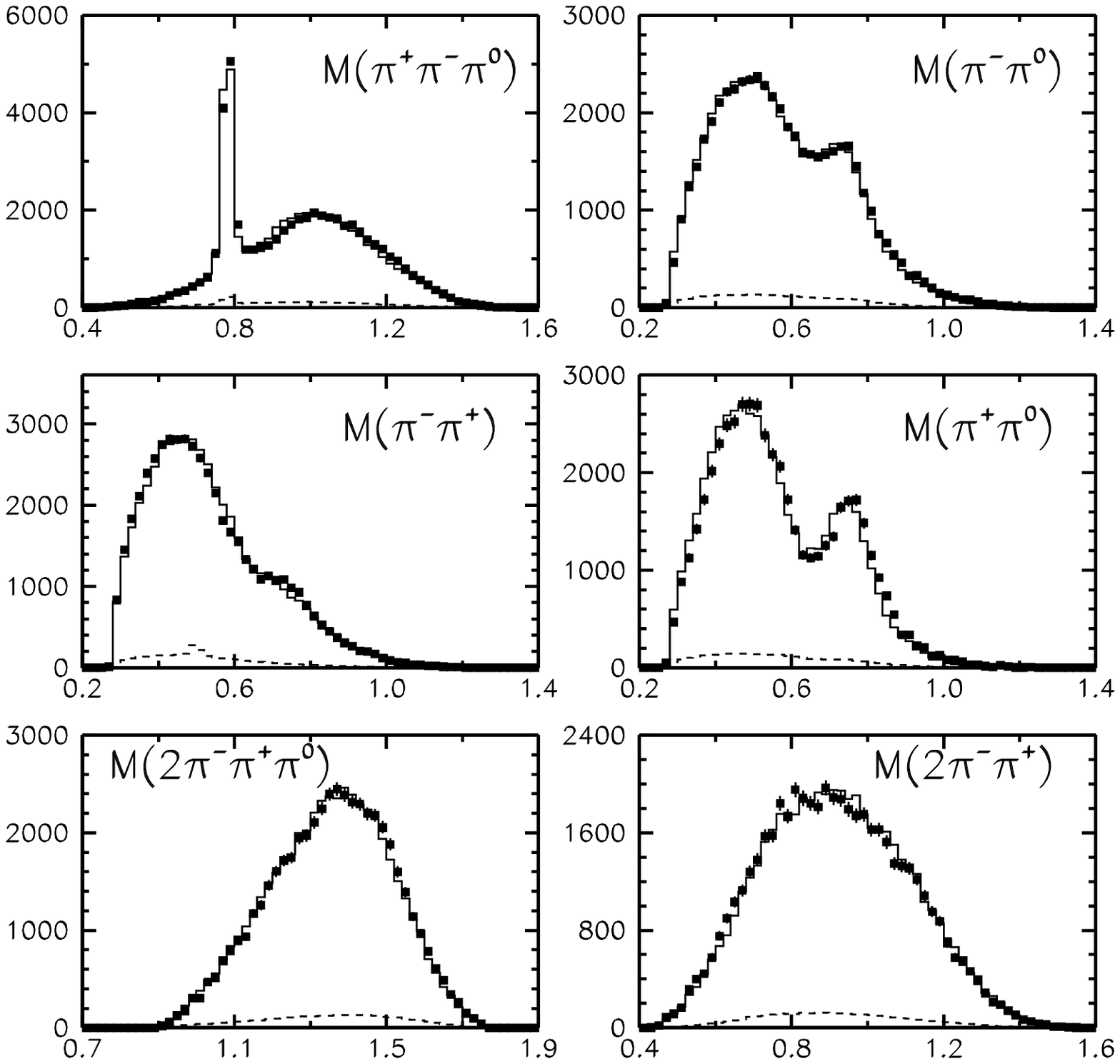}{fits_1}
{Comparison between fits and data distributions for the 
$\omega\pi$,
$\rho\pi\pi$ and non-resonant $3\pi\piz$ model. 
The horizontal axis is
given in units of $GeV/c^2$. Solid squares represent the data, solid
line represents the nominal fit, and the dashed line represents the
parameterized background.}

\simplex{htbp}{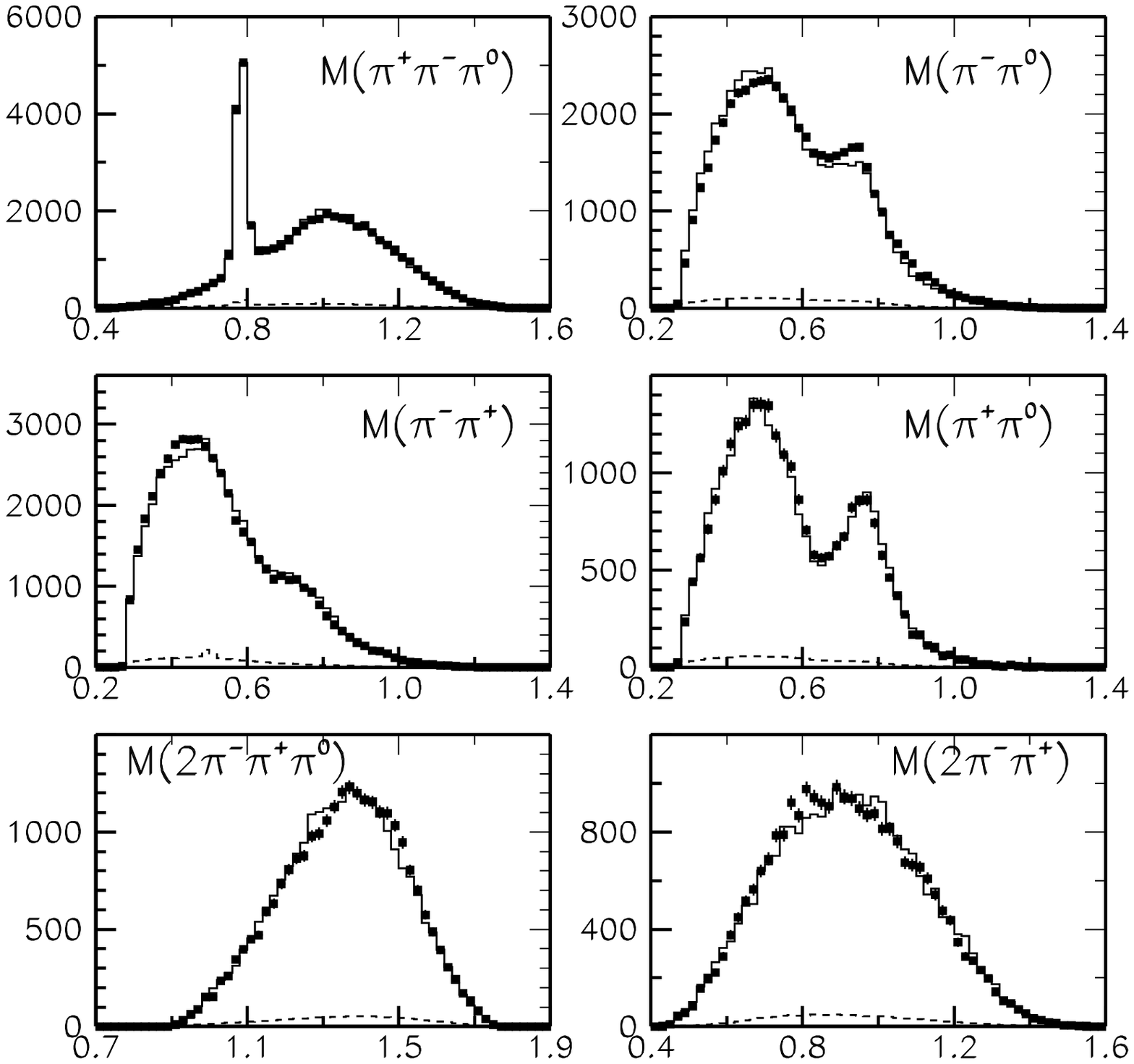}{fits_2}
{Comparison between fits and data distributions for the 
$\omega\pi$ and $a_1\pi$ model. The horizontal axis is
given in units of $GeV/c^2$. Solid squares represent the data, solid
line represents the nominal fit, and the dashed line represents the
parameterized background.}

\simplex{htbp}{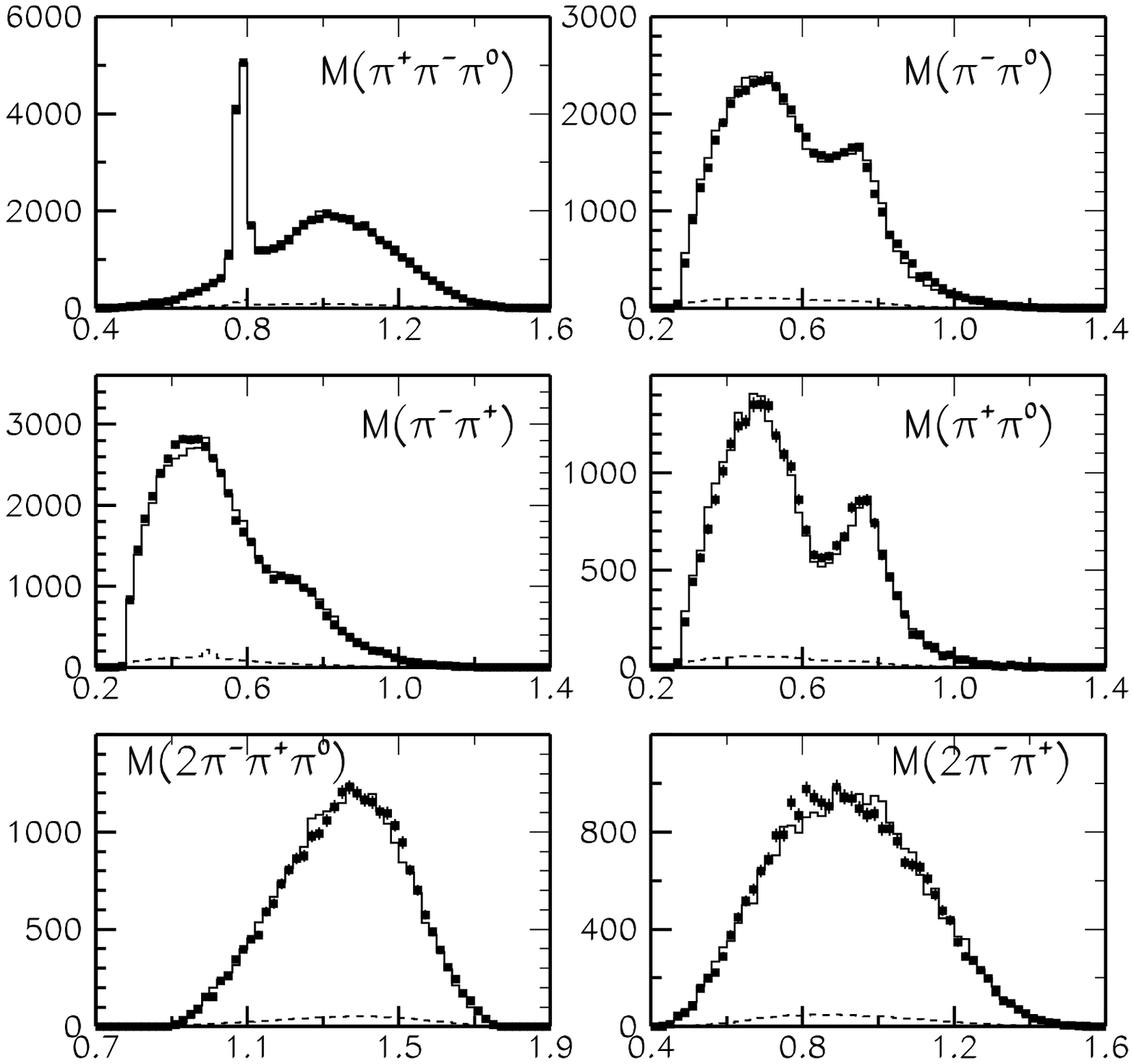}{fits_3}
{Comparison between fits and data distributions for the 
$\omega\pi$, 
$a_1\pi$, $\sigma\rho$ and $f_0\rho$ model. The horizontal axis is
given in units of $GeV/c^2$. Solid squares represent the data, solid
line represents the nominal fit, and the dashed line represents the
parameterized background.}

\subsection{Systematic errors}

 To estimate systematic biases and errors for every model,
we generate 40 Monte Carlo samples with amplitudes obtained from 
the fit to the data and then fit them to the chosen parameterization of
the hadronic current. Each sample is mixed with simulated 
background events in the proportion expected from the data.  We integrate 
the results of each simulation to estimate 
necessary corrections and systematic
errors for each of the amplitudes.  Biases of the reconstructed
amplitude values are shown in Table~\ref{tab:results}.
Errors are determined as root mean squares of the 
parameter fluctutations. 

The total systematic error also includes uncertainty due to the choice
of the parameters of $\rho'$.
We use PDG~\cite{PDG} values for the masses 
and widths of the $\rho$ and $\rho''$ mesons and vary the $\rho'$
parameters within the errors obtained in the fit described in
Section~\ref{sec:ompi_fit}. 
We also switch to a smaller mass of $\rho'$~\cite{tau96}:
$M_{\rho'}=1370\ \mbox{MeV}/c^2$. 
The results appear to have very little sensitivity to the choice of
the parameters for $\rho'$.

All contributions to the systematic error are added in quadrature.
Systematic biases for the components of the hadronic current are
shown in Table~\ref{tab:systematics}.

\begin{table}[bthp]
\caption{}{\label{tab:systematics}
Systematic biases for the hadronic current components.
Systematic bias is defined as
$\delta_i = (R_i^{reconstructed} -R_i^{generated}) / R_i^{generated}$.}
\begin{center}
\begin{tabular}{|c|c|} \hline
Model 		& Systematic bias $\delta_i$ \\ \hline
&  $\delta_{\rho^0\pi^-\piz}=(+4.0\pm 3.8)\%$ \\
&  $\delta_{\rho^-\pi^-\pi^+}=(-7.6\pm 3.8)\%$ \\
Model 1
&  $\delta_{\rho^+\pi^-\pi^-}=(+2.3\pm 2.6)\%$ \\
&  $\delta_{\omega\pi}=(+2.5\pm 2.6)\%$ \\
&  $\delta_{3\pi\piz}=(+1.2\pm 9.0)\%$ \\ \hline
Model 2 
&  $\delta_{\omega\pi}=(+1.0\pm2.5)\%$ \\
&  $\delta_{a_1\pi}=(-2.0\pm2.5)\%$ \\ \hline
&  $\delta_{\omega\pi}=(+1.0\pm2.5)\%$ \\
Model 3
&  $\delta_{a_1\pi}=(-2.0\pm2.6)\%$ \\
&  $\delta_{\sigma\rho}=(0.0\pm 14.0)\%$ \\
&  $\delta_{f_0\rho}=(+1.0\pm 16.0)\%$ \\ \hline
& $\delta_{\omega\pi}=(+1.0\pm2.5)\%$ \\
& $\delta_{a_1\pi}=(-2.0\pm2.5)\%$ \\
Model 4
& $\delta_{\rho^0\pi^-\piz}=(1.0\pm 23.0)\%$ \\
& $\delta_{\rho^-\pi^-\pi^+}=(-0.0\pm 15.0)\%$ \\
& $\delta_{\rho^+\pi^-\pi^-}=(-2.5\pm 33.0)\%$ \\ \hline
\end{tabular}
\end{center}
\end{table}

\subsection{Comparison with similar analyses}

A recent analysis by the ALEPH Collaboration~\cite{aleph} assumes 
$\omega\pi$ and non-resonant $\rho\pi\pi$ contributions
and uses a simplified fit to the
submass projections without assigning systematic errors. 
This analysis by ALEPH gives the
following amplitudes: $R_{\rho^0\pi^-\piz}=0.11$, 
$R_{\rho^-\pi^-\pi^+}=0.20$, 
$R_{\rho^+\pi^-\pi^-}=0.22$, and $R_{\omega\pi}=0.40$.
Predictions from the chiral perturbation theory~\cite{chiral} for
the same model give: 
$R_{\rho^-\pi^-\pi^+}=0$, 
$R_{\rho^+\pi^-\pi^-}/R_{\rho^0\pi^-\piz}=2$. 
These results can be also compared with the ``standard mix''
expectations:
$R_{\rho^0\pi^-\piz}=0.28$, $R_{\rho^-\pi^-\pi^+}=0.27$, 
$R_{\rho^+\pi^-\pi^-}=0.08$, and $R_{\omega\pi}=0.30$.

\subsection{CVC test}

The CVC hypothesis relates the weak charged hadronic current in the
$\tau$ lepton decay 
to the isovector part of the electromagnetic current. 
Thus, the spectral function~(\ref{eq:spec_ff}) 
can be related~\cite{tsai} 
to the cross-section for $e^+e^-\to 4\pi$ as:
\begin{equation}
V^{3\pi\piz}(q) = \frac{q^2}{4\pi^2\alpha^2}
\sigma_{e^+e^- \to 4\pi}(q)\ .
\end{equation}
Therefore, the
resonant structure 
of the hadronic final state in the decay $\tau\to 3\pi\piz\nu_\tau$ 
should be comparable to that in the
reactions $e^+e^-\to 2\pi^+ 2\pi^-$ and
$e^+e^-\to\pi^+\pi^- 2\pi^0$  
in the energy range corresponding to the $\tau$ mass.
The only expected difference is due to the fact that the $\omega\pi$ final
state is not accessible  
in the all charged final state. 
A recent analysis of the CMD-2 data~\cite{Novosibirsk} shows
clear dominance of the $a_1\pi$ component
in the four charged pion final state and of the 
$a_1\pi$ together with the $\omega\pi$ components 
in the 2 charged and 2 neutral pion mode. 

A plot comparing the $3\pi\piz$ and $\omega\pi$ spectral functions
in the decay $\tau\to 3\pi\piz\nu_\tau$ with the corresponding
cross-sections measured by the CMD-2 group~\cite{CMD2} is shown in 
Fig.~\ref{fig:spec_and_sigma}.
Our results are in excellent qualitative agreement and support the 
CVC hypothesis. 
\simplex{htbp}{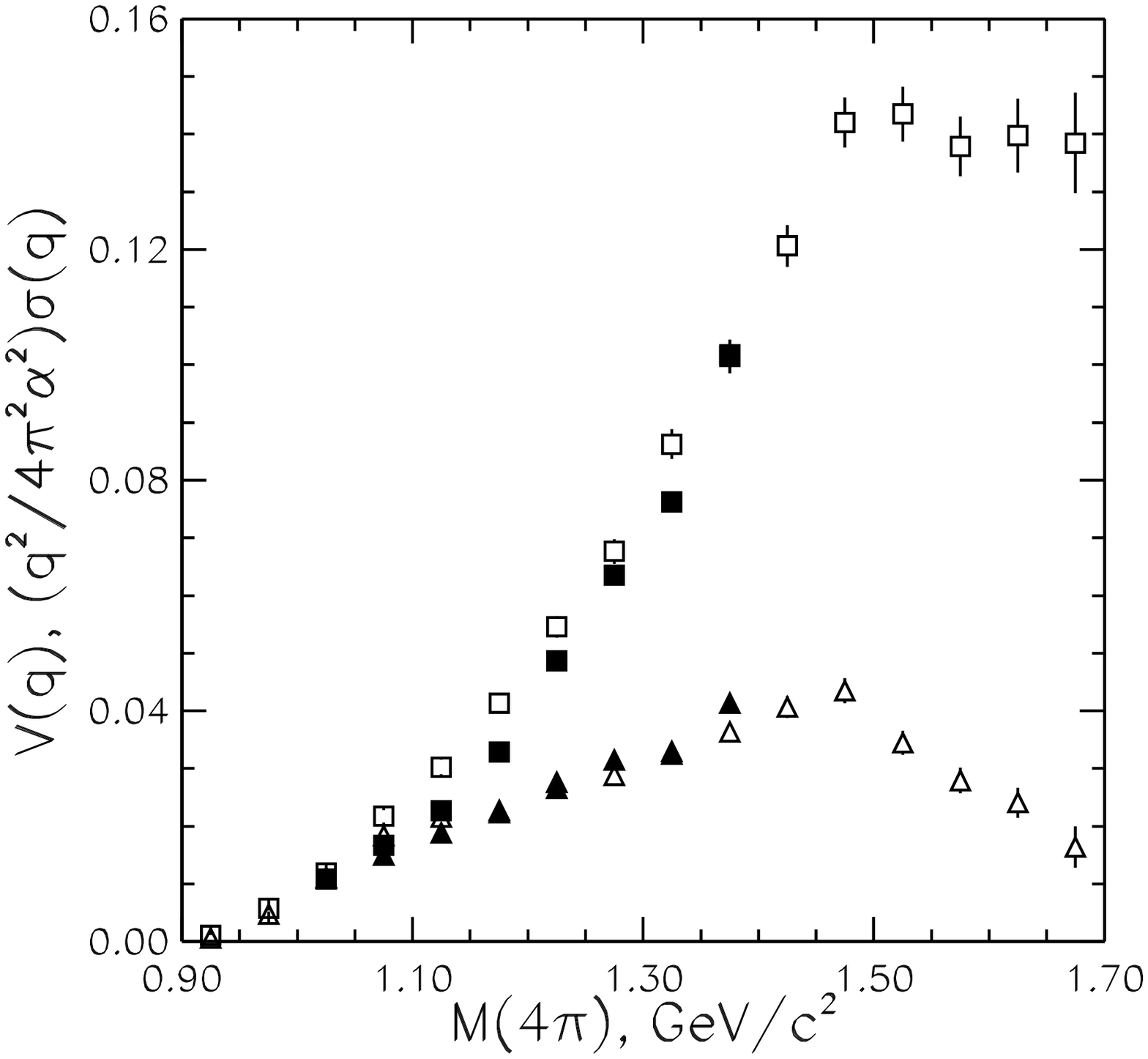}{spec_and_sigma}
{Comparison of $V^{3\pi\piz}(q)$ and $V^{\omega\pi}(q)$ measured in
this analysis with the corresponding cross-sections measured by
CMD-2. Squares represent the full four-pion spectrum, and triangles
represent the $\omega\pi$ component. Our results are shown with
open symbols, and the results obtained by CMD-2 are shown with filled
symbols.}

\section{Conclusions}

The large data sample of $\tau\to 3\pi\piz\nu_\tau$ decays collected by the
CLEO II experiment allowed for a detailed study of the resonance decomposition
and structure functions of the $3\pi\piz$ final state. The analysis indicates 
dominance of the $\omega\pi$ and $a_1\pi$ contributions 
which is in good agreement with
theoretical expectations based on the CVC hypothesis and 
Chiral Langrangian calculations. 
At the same time, 
with the present statistics, other models can provide satisfactory
descriptions of the existing resonant structure. 
The goodness-of-fit is slightly higher for the models with a dominating
$a_1\pi$ contribution, though the $a_1\pi$ channel alone does not
reproduce the observed non-$\omega$ contribution precisely.
The mass and width of the $\rho'$ resonance have been extracted from
the fit to the $\tau\to\omega\pi\nu_\tau$ spectral function,
and a new upper limit on the non-vector
current contribution to this spectral function of 5.4\% 
at 90\% CL has been obtained.

\section{Acknowledgments}
We gratefully acknowledge the effort of the CESR staff in providing us with
excellent luminosity and running conditions.
J.R. Patterson and I.P.J. Shipsey thank the NYI program of the NSF, 
M. Selen thanks the PFF program of the NSF, 
M. Selen and H. Yamamoto thank the OJI program of DOE, 
J.R. Patterson, K. Honscheid, M. Selen and V. Sharma 
thank the A.P. Sloan Foundation, 
M. Selen and V. Sharma thank the Research Corporation, 
F. Blanc thanks the Swiss National Science Foundation, 
and H. Schwarthoff and E. von Toerne thank 
the Alexander von Humboldt Stiftung for support.  
This work was supported by the National Science Foundation, the
U.S. Department of Energy, and the Natural Sciences and Engineering Research 
Council of Canada.

\newpage
\appendix
\section{Hadronic current for the decays 
$\tau\to a_1\pi\nu_\tau$,
$\tau\to\sigma\rho\nu_\tau$ and $\tau\to f_0\rho\nu_\tau$}

Expressions for the hadronic current for $\tau$ decays into the four
pion final state via $a_1\pi$,
$\sigma\rho$ and $f_0\rho$ channels are derived under the assumption that
the $\tau$ decays into an intermediate vector resonance $\tilde{\rho}$:
$\tau\to\tilde{\rho}\nu_\tau$, which
consecutively decays into the shown channels. 
The vector boson $\tilde{\rho}$ has quantum numbers $I(J^{PC})=1(1^{--})$, 
and its 
spectral function, in accordance with Eqn.~(\ref{eq:fbw}),
is simulated as mixture of $\rho$ and $\rho'$
resonances and a phase space contribution.
Assuming isospin conservation and enforcing Bose symmetry, one obtains
the following expressions for the form factors $f_i^\mu$ 
in the decay $\tau^-\to 2\pi^-\pi^+\piz\nu_\tau$:
\begin{eqnarray}
f_{a_1\pi}^\mu & = & T^{\mu\nu} (q) \times \nonumber\\
      &   &  \bigg[   \phantom{-}           \,\,
            T_{\nu\kappa} (Q_1 )     BW_{a_1} ( Q^2_1 ) 
             \left[ BW_\rho ( ( q_1 + q_2 )^2 ) ( q_1 - q_2 )^\kappa + 
                    BW_\rho ( ( q_1 + q_4 )^2 ) ( q_1 - q_4 )^\kappa  \right] 
\nonumber\\
      &   & \phantom{\bigg[}
            - T_{\nu\kappa} (Q_2 )     BW_{a_1} ( Q^2_2 ) 
             \left[ BW_\rho ( ( q_1 + q_3 )^2 ) ( q_1 - q_3 )^\kappa + 
                    BW_\rho ( ( q_2 + q_3 )^2 ) ( q_2 - q_3 )^\kappa  \right] 
\nonumber\\
      &   & \phantom{\bigg[} 
            - T_{\nu\kappa} (Q_3 )     BW_{a_1} ( Q^2_3 ) 
             \left[ BW_\rho ( ( q_1 + q_3 )^2 ) ( q_1 - q_3 )^\kappa + 
                    BW_\rho ( ( q_3 + q_4 )^2 ) ( q_4 - q_3 )^\kappa  \right] 
            \bigg] \ ; 
\end{eqnarray}
\begin{eqnarray}
\label{eq:sigma_rho}
f_{\sigma\rho}^\mu & = & T^{\mu\nu} (q) \times
             \bigg[\phantom{+}  BW_\sigma ( ( q_1 + q_2  )^2 )  
BW_\rho ( ( q_3 + q_4 )^2 ) 
                    ( q_4 - q_3 )_\nu   \nonumber\\
      &   &  \qquad\qquad\qquad \qquad\,\,  
                              + BW_\sigma ( ( q_1 + q_4  )^2 )  
BW_\rho ( ( q_2 + q_3 )^2 ) ( q_2 - q_3 )_\nu \, \bigg] \ ;
\end{eqnarray}
where
\begin{eqnarray}
q_1 & : = & \mbox{ four-momentum of } \pi^+ \ ;   \nonumber\\
q_2 & : = & \mbox{ four-momentum of } \pi^-_1 \ ; \nonumber\\
q_3 & : = & \mbox{ four-momentum of } \pi^0 \ ;   \\
q_4 & : = & \mbox{ four-momentum of } \pi^-_2 \ ; \nonumber
\end{eqnarray}
\begin{eqnarray}
q   & = &  q_1 + q_2 + q_3 + q_4 \ ; \nonumber\\
Q_1 & = & q_1 + q_2 + q_4 \ ; \nonumber\\
Q_2 & = & q_1 + q_2 + q_3 \ ; \\
Q_3 & = & q_1 + q_3 + q_4 \ ; \nonumber 
\end{eqnarray}
and $T^{\mu\nu} (x)$ denotes the projection operator:
\begin{equation}
T^{\mu\nu} (x) = -g^{\mu\nu} + \frac{x^\mu x^\nu}{x^2}\ .
\end{equation}
The expression for the $\tau$ decay into four pions via the $f_0\rho$
channel is identical to
Eqn.~(\ref{eq:sigma_rho}) with the mass of $f_0$ substituted instead
of the mass of $\sigma$. 

\end{document}